\documentclass[11pt,twoside,a4paper]{article}
\usepackage{graphicx}
\usepackage[english]{babel}
\usepackage[utf8] {inputenc}
\usepackage{amsmath,amsfonts,amssymb}
\usepackage{bbm}
\usepackage{stmaryrd}
\usepackage[all]{xy}
\usepackage{fancyhdr}

\newcommand\be{\begin{equation}} \newcommand\ee{\end{equation}}
\newcommand\bd{\begin{displaymath}} \newcommand\ed{\end{displaymath}}
\renewcommand\r{\mathbb{R}}
\newcommand\K{\mathcal{K}}
\newcommand\Kbar{\mathcal{\bar{K}}}
\renewcommand\L{\mathcal{L}}

\renewcommand\P{\mathcal{P}}
\newcommand\e{\varepsilon}
\renewcommand\l{\lambda}

\begin{document}

\title{ OPTIMAL INVESTMENT AND PRICE DEPENDENCE IN A SEMI-STATIC MARKET }

\author{Pietro Siorpaes }

\maketitle

\begin{abstract}
This paper studies the problem of maximizing expected utility from terminal wealth in a semi-static market composed of derivative securities, which we assume can be traded only at time zero, and of stocks, which can be traded continuously in time and are modeled as locally-bounded semi-martingales.
 Using a general utility function defined on the positive real line, we first study existence and uniqueness of the solution, and then we consider the dependence  of the outputs of the utility maximization problem on the price of the derivatives, investigating not only stability but also differentiability, monotonicity, convexity and limiting  properties.
\end{abstract}

\section{Introduction}
\label{Introduction}
   
A classical problem in financial economics is to understand the behavior of rational agents faced with an uncertain evolution of asset prices. 
In one of the most popular frameworks, one considers an investor who
wants to maximize his expected utility from terminal wealth by investing in a frictionless market.
 This approach takes as inputs a utility function and a model for the future evolution of the stock prices;
 to implement this program in practice, typically one chooses a particular parametrized family of utility functions and of models for the stock price,
and calibrates the value of the parameters to the available data.

Since the choice of the agent's utility and of the market model requires estimation, 
it is natural to ask how the agent’s behavior is affected by misspecifications of  the utility function and of the underlying market model.
Indeed, following Hadamard's prescription, after investigating existence and uniqueness 
one should perform stability analysis, and only a problem whose solution exists, is unique and depends continuously on the initial data is worthy of the appellative `well posed'.

In fact, much work has recently been done on the topic of sensitivity of the solution of the problem of expected utility maximization under perturbations of various initial conditions.
 Jouini and Napp \cite{JoNa04}, Carasus and Rásonyi \cite{CaRa07}, and Larsen \cite{La09} consider misspecifications of the utility functions. 
Market perturbations are considered in Larsen and {\v{Z}}itković \cite{LaZi07} who, working in a continuous time market, investigate the continuous dependence  on the price of the stock (parametrized by the market-price of risk), and in Kardaras and {\v{Z}}itković \cite{KaZi11}, who perform a stability analysis of the problem under small misspecifications of the agent's preferences and of the market model. 
Hubalek and Schachermayer \cite{HuSc98} study the convergence of prices of illiquid assets when the prices of the liquid assets converge, and stability of option pricing under market perturbations has been investigated by El Karoui et al.  \cite{KaJeSh98}, while Kardaras \cite{Kar09} looks at the stability of the numéraire portfolio.

The previously mentioned results only deal with stability, and constitute a zeroth order approach to the problem.
References which perform a first-order study are Henderson \cite{Hend:02}, who 
studies, in a Brownian market,  the expansion of the indifference price 
 with respect to a small number of random endowments (see also Henderson and Hobson \cite{HendHobs:02}), and Kramkov and Sirbu  \cite{KramSirb:06b}, who generalize the first order approximation to  semimartingale markets.

A setting which has been popular in recent years is the one of a semi-static market, where investors can trade continuously in time  a number of stocks, as well as take static positions in some derivatives; for example see  Campi \cite{Ca04,Ca2011ve},  Ilhan et. al \cite{Ilhan:06fj,IlJoSir:05} and Carr et al. \cite{Carr:2001vn},  Schweizer and Wissel \cite{Sch:2008zr,Sch:2008ys}, Jacod and Protter \cite{Jacod:2010ly}.
On of the advantages of this framework is that the price of a contingent claim which  can be traded  only at time zero is modeled simply as a vector in $\r^n$, instead of a general $\r^n$-valued semimartingale.
Working in this simplified setting and using an exponential utility, Ilhan et al. \cite{IlJoSir:05}  obtain \emph{differentiability and strict-convexity} of the value function  \emph{as a function of the price} of the financial derivatives.

In the present paper we consider, as Ilhan et al. \cite{IlJoSir:05},  the problem of maximizing expected utility from terminal wealth  in a semi-static market  framework; however, we use a general utility function defined on the positive real line.
We study the existence and uniqueness of the solution, and the dependence of the value function, of its maximizer and of other quantities of interest on the (initial capital and on the) price $p$ of the derivatives; we  prove not only stability, but also  differentiability, monotonicity, and convexity. Specifically, we reproduce in our model some economically sensible properties: the value function $u$ has the expected monotonic behavior (as $p$ increases $u$ is initially decreasing, then constant, then increasing), and it diverges (along with the optimizer) when the price of the derivatives converge to an arbitrage price. 
We also show  convexity in $p$ of  the largest feasible position, defined as the maximum number of shares of derivatives that the agent with given initial wealth can buy at price $p$ and still be able to invest in the (liquid) stock market as to have a non-negative final wealth. This fact, which is a consequence of the no-arbitrage assumption, is in our opinion not particularly intuitive (especially when there are multiple derivatives) and -to the best of our knowledge- was not noticed before.
Unlike in the case of exponential utility, we show with an example that the maximal expected utility does not need to be a convex function of the derivative's prices; this makes proving differentiability a much trickier task, which we are able to carry on only under additional assumptions  (most importantly in the case of power utility).

We emphasize that our problem does not fall under the general umbrella of utility maximization with convex constraints (we refer to  Larsen and {\v{Z}}itkovi{\'c}  \cite{LarZit13a} for a survey), since it cannot be re-phrased asking that the portfolio and wealth process lie in some given convex set (possibly depending on $t$ and $\omega$); rather, we demand that the investor, after choosing his position at time zero \emph{arbitrarily}, keep his position in derivatives unchanged for the rest of the time horizon, while freely investing in stocks.
Moreover, Ilhan et al. \cite{IlJoSir:05}, working in the framework of exponential utility, make essential use of relative entropy techniques and of  some explicit representations of the maximal utility and of indifference prices; thus, as they point out, the `extension to more general cases is not trivial', as a completely different approach is needed.

 A simple but useful observation is that our problem can be decomposed in two steps: 
choosing the optimal amount of derivatives to buy at time zero, and then investing optimally in the continuous time stock market; the second step being the problem of optimal investment with random endowment, which is then closely related to our problem.
To describe this relationship and make profitable use of it, we will need to slightly extend the main result of Hugonnier and Kramkov  \cite{HugonKram:04} by considering endowments on the boundary of the domain of the utility;  after personal communications with us, Mostovyi, who had generalized the results of \cite{HugonKram:04}  to include the case of intermediate consumption, has  analogously extended  in  \cite{Most11Op}  his results  to include endowments on the boundary.

The  paper is organized as follows. In Section \ref{The model} we present
the model of financial market and we define our problem. In Section  \ref{Statement of the main theorem}
we state our main theorems, some of which we state in more detail  in Sections \ref{Existence and Uniqueness} and \ref{Continuity}.
 In Section \ref{Optimal investment with stocks and derivatives} we prove existence and uniqueness of the solution. In Section \ref{Optimal investment with random endowment} we generalize some results of Hugonnier and Kramkov \cite{HugonKram:04}, and in Section \ref{Relation between the two optimization problems} we 
study the relation between our problem and the problem of optimal investment with random endowment.
 In Section \ref{Continuity} we establish the continuity of the outputs of the utility maximization problem. In Section \ref{asymptotic} we prove the convexity of the largest feasible position and we study the asymptotic behavior of the value function and the optimizer. In Section \ref{1dc} we consider in more detail the one-dimensional case, and in Section \ref{non-convex} we provide an example of a value function which is not convex  in $p$. Finally, in Section \ref{Differentiability} we study the differentiability of the value function.

\section{The model}
     \label{The model}
Consider at first a model of a financial market composed of a savings
account and $d$ stocks which can be traded continuously in time.  As
is common in mathematical finance, we consider a finite deterministic\footnote{In fact, one could take $T$ to be a finite stopping time, as is the case in Hugonnier and Kramkov (2004), on which we rely.}  time horizon
$[0,T]$, and we assume that the interest rate is $0$; that is, the
price process of the savings account is used as numéraire and is thus
normalized to one.  The price process $S = (S^i)_{i=1}^d$ of the
stocks is assumed to be a locally-bounded\footnote{This assumption is not strictly speaking necessary, as the results in \cite{KramSch:99}, \cite{KramSch:03}, \cite{HugonKram:04},  \cite{DelbSch:97} on which our proofs hinge, although proved for a locally-bounded semi-martingale, are true also without the local boundedness assumption, if one replaces equivalent local-martingale measures with separating measures throughout (this fact is stated in \cite[Remark 3.4]{HugonKramSch:05}).}
 semi-martingale on a given filtered
probability space $(\Omega,\mathcal{F},\mathbb{F},P)$, where the
filtration $\mathbb{F}=(\mathcal{F}_t)_{t\in [0,T]}$ satisfies the
usual conditions.
 
Now enlarge the market by allowing also $n$ contingent claims
$f=(f_j)_{j=1}^n$ to be traded at price $p=(p_j)_{j=1}^n$; we assume
that these contingent claims can be traded \emph{only} at time zero, and that $p$ is an arbitrage-free price for the European contingent claims $f$ (in a sense which will be made precise later).
  
A self-financing portfolio is then defined as a triple $(x,q,H)$,
where $x\in \r$ represents the initial capital, $q_j\in \r$ represents
the holding in the contingent claim $f_j$, and the random variable
$H_t^i$ specifies the number of shares of stock $i$ held in the
portfolio at time $t$.

An agent with portfolio $(x,q,H)$ will invest his initial wealth $x$
buying $q$ European contingent claims at price $p$ at time zero. This
quantity is then held constant up to maturity, so the vector $q$
represents the illiquid part of the portfolio and $q p:=\sum_{j=1}^n
q_jp_j$ represents the wealth invested in the European contingent
claims (in this paper $vw$ will always denote the dot product between two vectors $v$ and $w$, and $|\cdot|$ will denote the Euclidean norm in $\r^{n+1}$).
  
He will then invest the remaining wealth $x-q p$ dynamically, buying
$H_t$ share of stocks at time $t \in [0,T]$, and put the rest (positive or
negative) into the savings account.  We will denote by $X_t$ the value
of the dynamic part of the portfolio, which will be called simply the
wealth process; $x-q p$ will be called the initial value of the wealth
process (which is different from the initial wealth $x$ of the
portfolio).  The wealth process $ X$ evolves in time as the stochastic
integral of $H$ with respect to $S$: \bd
\label{wealth} \textstyle
X_t=x-qp+(H\cdot S)_t=x-qp+\int_0^t H_u dS_u , \qquad t\in [0,T] , \ed
where $H$ is assumed to be a predictable $S$-integrable process.

For $x\geq0$, we denote by $\mathcal{X}(x)$ the set of non-negative
wealth processes whose initial value is equal to $x$, that is, \bd
\mathcal{X}(x):=\{X\geq 0 : X_t=x+(H\cdot S)_t \}.  \ed
	
A probability measure $Q$ is called an \emph{equivalent local-martingale
measure} if it is equivalent to $P$, and if $S$ is a local-martingale under $Q$.
 We denote by $\mathcal{M}$ the family of
equivalent local-martingale measures, and we assume that 
\be
\label{na}
\mathcal{M}\ne \emptyset .
  \ee
 This condition is essentially
equivalent to the absence of arbitrage opportunities in the market
without the European contingent claims: see 
\cite{DelbSch:94} and \cite{DelbSch:98} for precise statements as well as for
further references.

In our model we consider an agent whose preferences are modeled via a utility function $U:(0,\infty)\rightarrow \r$, which is assumed to be
strictly concave, strictly increasing and continuously differentiable
and to satisfy the Inada conditions: \be \textstyle
\label{inada}
U'(0):=\lim_{x \to 0+}U(x)=\infty , \qquad U'(\infty):=\lim_{x \to
  \infty}U'(x)=0 .  \ee It will be convenient to consider $U$  defined
on the whole real line. We want its extension to be concave and upper
semi-continuous, and (\ref{inada}) implies that there is only one
possible choice: we define $U(x)$ to be $-\infty$ for $x$ in
$(-\infty,0)$, and to equal $U(0+):=\lim_{x \to 0} U(x)$ at $x=0$.

We denote by $f=(f_j)_{j=1}^n$ the family of the
$\mathcal{F}_T$-measurable payment functions of the European
contingent claims with maturity $T$, and by $ qf=\sum_{j=1}^n q_j
f_j $ the payoff of the static part of the portfolio.  The total
payoff of the portfolio $(x,q,H)$ is then $x-qp+(H\cdot S)_T+qf$.

A non-negative wealth process in $\mathcal{X}(x)$ is said to be maximal
if its terminal value cannot be dominated by that of any other process
in $\mathcal{X}(x)$. 
We assume that the European contingent claims can be sub- and super-replicated by trading in the stock; in other words, that there exists a maximal wealth process $X'$ such that 
\be
\label{domination} \textstyle
|f|:=\sqrt{\sum_{j=1}^n f_j^2 } \leq X'_T \, .  \ee

Since  strictly positive maximal wealth processes are  numéraires (see    \cite{DelbSch:95}),  condition \eqref{domination} can be interpreted as asking that $|f|$ be bounded with respect to some numéraire.
If the contingent claims are uniformly bounded, as in \cite{CvitSchWang:01}, then the optimization set is taken to be the set of wealth processes uniformly bounded from below. If
the contingent claims are bounded just with respect to some numéraire,
the optimization set has to be extended analogously.
Following \cite{DelbSch:97},  we say that a wealth process $X$ is \emph{acceptable} if it admits a
representation of the form $X = X' - X''$, where $X'$ is a non-negative
wealth process and $X''$ is a maximal wealth process. 
 Since  a wealth process $X$ is  \emph{acceptable} if and only if it is  bounded below under some numéraire\footnote{If $X = X' - X''$ take $N=1+X''$  to get $X/N \geq -1$; vice versa if $X/N \geq -1$ choose $X''=N$ to get $X = X' - X''$.} $N$,  the  acceptable processes constitutes a natural optimization set for our optimal investment problem. 
Thus, following \cite{HugonKram:04}, we define $\mathcal{X}(x,q)$ to be the
set of acceptable wealth processes with initial value $x$ whose
terminal value dominates the random payoff $-qf$, i.e.,  \bd
\mathcal{X}(x,q):=\{ X : X \textrm{ is acceptable}, X_0 = x \textrm{
  and } X_T+ qf \geq 0 \} .  \ed
  We will call
$\mathcal{\bar{K}}$ the set of points $(x, q)$ where
$\mathcal{X}(x,q)$ is not empty, i.e.,  \be
\label{def of K}
\mathcal{\bar{K}}:=\{(x,q)\in \r \times \r^{n}: \mathcal{X}(x,q)\ne \emptyset
\} . \ee
As shown in \cite[Lemma 1 and 6]{HugonKram:04}, assumptions (\ref{na}) and (\ref{domination})  imply that  the convex cone $\mathcal{\bar{K}}$ defined in (\ref{def of K}) is
 closed and its interior $\mathcal{K}$ contains $(x,0)$ for any $x>0$,
 so $\mathcal{\bar{K}}$ is the closure of $\mathcal{K}$. 

We will say that $p$ is an \emph{arbitrage-free price} for the European contingent claims $f$ if any portfolio with
zero initial capital and non-negative final wealth has identically zero final wealth, and we will denote by $\P$ the set of arbitrage-free prices; in other words, 
 $$\P:=\{ p \in \r^n :  q\in\r^n, X\in\mathcal{X}(-pq,q) \text{ imply } X_T=-qf\}.$$

The objective of this paper is to study  the problem of utility maximization in the enlarged market consisting of the bond, the stocks and the contingent claims, i.e., the following optimization 
problem, for $ x>0, p\in \P$,
 \be
\label{prrendtilde}
\tilde{u}(x,p):=\sup{ \{ \mathbb{E}[U(X_T+qf)] : X \textrm{ is
    acceptable }, q\in \r^n , \, X_0 = x-qp\} }. \ee
We follow the convention that $\mathbb{E}[U(X_T+qf)]$ equals $-\infty$ when $\mathbb{E}[U^{-}(X_T+qf)]=-\infty$ (whether or not $\mathbb{E}[U^{+}(X_T+qf)]$ is finite).

The problem of utility maximization with
 random endowment\footnote{We use the convention that the $\sup$ (inf) over an  empty set takes the value $-\infty$ ($+\infty$).}
 \be
 \label{prrend} 
 u(x,q):=\sup_{X \in \,\, \mathcal{X}(x,q)} \mathbb{E}[U(X_T+qf)] ,
 \quad (x,q)\in \r^{n+1}, \ee
which was studied in \cite{HugonKram:04}, is obviously closely related to (\ref{prrendtilde}); in fact it is easy to show that $\mathcal{K} \subseteq \{u>-\infty\} \subseteq \mathcal{\bar{K}}$, and trivially
\be
 \label{equalus}
 \tilde{u}(x,p)=\sup_{q\in \r^n} u(x-qp,q)= \sup_{q\in \r^n : u(x-qp,q)> -\infty  } u(x-qp,q), \quad x>0, p\in \P.
\ee

 By definition, $p$ is a \emph{marginal (utility-based) price} at $(x,q)$ for  $f$  if the agent with initial endowment $(x,q)$ given the opportunity to trade the contingent claims $f$ at time zero at price $p$ would neither buy nor sell any. We will denote by  $\mathcal{P}(x,q)$ the set of marginal prices of $f$  at $(x,q)\in \{ u\in \r\}$; in other words, we set
 $$ \mathcal{P}(x,q):=\{p\in \r^n :  u(x-q'p,q+q') \leq u(x,q)  \text{ for all } q' \in \r^{n}\}.  $$
The \emph{largest feasible position}  $m:(0,\infty)\times \mathcal{P} \longrightarrow  [0,\infty]$ is defined as \be
 \label{m} \textstyle
 m(x,p):=\sup \left\{ \, |q| : q \in \r^n, (x-qp,q) \in \mathcal{\bar{K}} \, \right\},
 \ee
and it  measures  the maximum number of shares of derivatives that the agent with wealth $x$ can buy at price $p$ and still be able to invest in the stock market as to have a non-negative final wealth.

In this paper we study problem (\ref{prrendtilde}) and  its relationship with problems (\ref{prrend}) and \eqref{equalus}. In particular, we investigate  existence and uniqueness of the solution, and the dependence on the initial wealth  $x$ \emph{and the initial price $p$} of the outputs of problem (\ref{prrendtilde}): the optimal position in   derivatives $ \tilde{q}$, the optimal final value of the dynamic part of the portfolio  $\tilde{X}_T$, the maximal expected utility $ \tilde{u}$ (and its derivatives), and the largest feasible position.

 \section{Statement of the main theorems}
\label{Statement of the main theorem}
To state our main theorems we need to introduce some standard notation.
 Denote by $\mathcal{Y}(y)$ the family of non-negative processes $Y$
 with initial value $y$ and such that for any non-negative wealth
 process $X$ the product $XY$ is a super-martingale, that is, \bd
 \mathcal{Y}(y):=\{Y\geq0 : Y_0 = y , XY \textrm{ is a super-martingale
   for all } X\in \mathcal{X}(1)\}.  \ed In particular, as
 $\mathcal{X}(1)$ contains the constant process $1$, the elements of
 $\mathcal{Y}(y)$ are non-negative super-martingales. Note also that the
 set $\mathcal{Y}(1)$ contains the density processes of all
 $Q\in\mathcal{M}$. 

 The convex conjugate function $V$ of the agent's utility function $U$
 is defined to be the Fenchel-Legendre transform of the function $-U(-
 \cdot)$; that is, \bd \textstyle V(y):=\sup_{x\in \r}(U(x)-xy)=\sup_{x>0}(U(x)-xy), \quad y\in \r.  \ed It
 is well known that, under the Inada conditions (\ref{inada}), the
 conjugate $V$ of $U$ is convex, lower semi-continuous, it is infinite on $(-\infty,0)$ and, restricted to $(0,\infty)$, is a continuously differentiable, strictly
 decreasing and strictly convex function satisfying $V'(0)= -\infty$,
 $V'(\infty)=0$, $V (0) = U(\infty)$, $V (\infty) = U(0)$, as well
 as the following bi-dual relation: 
\bd
\textstyle U(x) = \inf_{y\in \r}(V (y) +xy)=\inf_{y>0}(V (y) +xy),
 \quad x \in \r.  \ed

 Following \cite{HugonKram:04}, we will denote by $w$ the value function of the problem of optimal
 investment without the European contingent claims, and by $\tilde{w}$
 its dual value function. In other words \bd w(x):=\sup_{X \in \,\,
   \mathcal{X}(x)} \mathbb{E}[U(X_T)] ,\quad x>0 \, ; \qquad
 \tilde{w}(y):=\inf_{Y\in\mathcal{Y}(y)}\mathbb{E}[V(Y_T)] , \quad
 y>0.  \ed

 Recall that a random variable $g$ is said to be \emph{replicable} if
 there is an acceptable process $X$ such that $-X$ is also acceptable
 and $X_T = g$ (if such a process exists, it is unique).
In order to have uniqueness of the maximizer of (\ref{equalus}) 
we will assume that \be
 \label{wlog}
 \textrm{ for any non-zero $q\in \r^n$ the random variable $qf$ is not
   replicable} .  \ee
Note that, by discarding the redundant contingent claims, one can always reduce to the case where (\ref{wlog}) holds (see  \cite[Remark 6]{HugonKram:04}), so assuming (\ref{wlog})  does not comport a real loss of generality.

The following theorem is an  analogue of the results found in \cite{KramSch:03}, plus a description of the relationship between problems (\ref{prrendtilde}), (\ref{prrend}), and (\ref{equalus}); we state it here in an abridged format (for the full version see Theorems \ref{th2} and  \ref{comparison}).

 \newtheorem{thmerge}{Theorem}
 \begin{thmerge}
   \label{thmerge}
   Assume that $p$ is an arbitrage-free price for $f$, that conditions
   (\ref{na}), (\ref{inada}), (\ref{domination}) and  (\ref{wlog}) hold, and
   that
 \bd
   \tilde{w}(y) <\infty \textrm{ for all } y>0.  
\ed

   Then the value function $ \tilde{u}(\cdot,p)$ is finite,
     continuously differentiable, strictly increasing and strictly
     concave on $(0,\infty)$ and satisfies Inada conditions.

For any $x > 0$, the solution $(\tilde{X},\tilde{q})$ to
  (\ref{prrendtilde}) and  the solution 
$(x-\hat{q}p,\hat{q})$ to (\ref{equalus})     exist, are unique and satisfy
$\tilde{q}= \hat{q}$.
Moreover, for every $(x,q) \in \{ u >-\infty \}$  the solution $X(x,q)$ to  (\ref{prrend})  exist, is unique and satisfies
$ \quad \tilde{X}= X(x-\hat{q}p,\hat{q}) .$
 \end{thmerge}

In Theorem \ref{thmerge}, the delicate point is that
in general\footnote{As we will show in Theorem \ref{th2}, a sufficient condition for  the solution of problem  (\ref{equalus}) to lie in $\K$ is that the final value of the optimal portfolio is bounded below by a strictly positive constant.} problem  (\ref{equalus}) does not have a solution
 if we were to replace $ \{ u >-\infty \}$  (or $\mathcal{\bar{K}}$) with its interior  $ \mathcal{K}$ (see \cite[Section 4]{Sio12:ArbFreePr}); however, in the existing literature problem (\ref{prrend}) has been solved only in the case where $(x,q)$ belongs to  $ \mathcal{K}$.
As a consequence, to compare the  problems (\ref{prrendtilde}) and (\ref{equalus}) and establish Theorem  \ref{thmerge},  we need to finish carrying out an extension of the results of   \cite{HugonKram:04} which was started in \cite{Sio12:ArbFreePr}.

 The following  theorem does not have an analogue in \cite{IlJoSir:05}
 and, we believe, is very intuitive (given its economic
 interpretation).  It shows that the dependence on $p$ takes a
 particularly pleasing form in the case where $q$ is one-dimensional.

 \newtheorem{1d}[thmerge]{Theorem}
 \begin{1d}
   \label{1d}
      Under the assumptions of Theorem \ref{thmerge}, if there is only one European contingent claim $f$, then there
   exists $\underline{p}, \overline{p}\in \r$ such that
   $(\underline{p},\overline{p})=  \mathcal{P}$
and, for all $x>0$, there   exists $a,b$ such that
   $\mathcal{P}(x,0)=[a,b]\subset (\underline{p},\overline{p})$.
 The function
   \begin{eqnarray}
     (\underline{p},\overline{p})& \longrightarrow & \r \nonumber\\
     p &\mapsto & \tilde{q}(x,p) \nonumber 
   \end{eqnarray}
defined in Theorem \ref{thmerge}   is continuous,  it is strictly positive on $(\underline{p},a)$,
   it equals zero on $[a,b]$, and it is strictly negative on
   $(b,\overline{p})$.  The function
   \begin{eqnarray}
     (\underline{p},\overline{p})& \longrightarrow & \r \nonumber\\
     p &\mapsto & \tilde{u}(x,p) \nonumber 
   \end{eqnarray}
   is continuous, it is strictly decreasing on
   $(\underline{p},a)$, it is constant on $[a,b]$, and it is strictly
   increasing on $(b,\overline{p})$.

 Moreover if $U(\infty)=\infty$ then
$\tilde{q}(x,\underline{p}+)=\tilde{u}(x,\underline{p}+)= \tilde{u}(x,\overline{p}-)=\infty$ and $\tilde{q}(x,\overline{p}-)=-\infty$.
 \end{1d}

 We recall that, in the general setting in which we work,
 $\mathcal{P}(x,0)$ is not a singleton (see \cite[Theorem
 3.1]{HugonKramSch:05}), so in the above theorem it could actually happen  that $a<b$.

 Our next theorem is a stability result for the problem of optimal investment in the general multi-dimensional setting (for a more detailed statement see Theorem  \ref{continuity3}).
It also shows that
$m$ is a convex function of the derivative's price and that, when the  arbitrage-free prices converge to an arbitrage price, the
 corresponding utility and the optimal demand diverge.
 \newtheorem{continuity1}[thmerge]{Theorem}
 \begin{continuity1}
   \label{continuity1}
      Under the assumptions of Theorem \ref{thmerge}, $\mathcal{P}$ is an open bounded convex set, the map $m(x,p)$ is finite valued, the  map
   \begin{eqnarray}
     (0,\infty) \times \mathcal{P} & \longrightarrow & \r^n \times
     L^0(P) \times \r \times \r \times \r  \nonumber\\
     (x,p) & \mapsto & \left( \tilde{q},\,\tilde{X}_T,\,
       \tilde{u},\,  \partial_x \tilde{u}, \,m      \right) \nonumber 
   \end{eqnarray}
   is continuous, $m(x,p)$ is positively homogeneous in $x$ and is convex in $p$, and the supremum in (\ref{m}) is attained.
 Moreover if $\mathcal{P} \ni p_n \to p \notin   \mathcal{P}, x_n \to x>0$ and $U(\infty)=\infty$ then, $ \text{ as } n \to \infty$, 
\bd
 \tilde{u}(x_n,p_n) \to \infty =   \tilde{u}(x ,p) , \quad
m(x_n,p_n) \to \infty \text{ and \, } 
 |\tilde{q} (x_n,p_n)| \to \infty .
   \ed
 \end{continuity1}

 We remark that in \cite{IlJoSir:05} the function $p \mapsto
 \tilde{u}(x,p)$ (corresponding to an exponential utility $U$) is
 proved to be strictly convex and differentiable. However, we will show that in our general framework convexity does not hold; this, and the fact  that the maximizer of 
 (\ref{equalus}) may lay on the boundary of $\mathcal{K}$  make proving differentiability in $p$ a very delicate task. The next theorem shows that these are the only impediments to differentiability, and that they can circumvented in some occasions.

 \newtheorem{udiff}[thmerge]{Theorem}
 \begin{udiff}
   \label{udiff}
   Under the assumptions of Theorem \ref{thmerge}, the function
   \begin{eqnarray}
     (0,\infty) \times \mathcal{P} & \longrightarrow & \r \nonumber\\
     (x,p) &\mapsto & \tilde{u}(x,p) \nonumber 
   \end{eqnarray}
   is continuously differentiable at all points
if the function $U$ is a power utility, and is continuously differentiable in a neighborhood of  some point $(x^{\star},p^{\star})$ if either of the following conditions is satisfied:
   \begin{enumerate} 
 \item The optimizer $(x^{\star}- \hat{q}(x^{\star},p^{\star}) p^{\star}, \hat{q}(x^{\star},p^{\star}))$ of \eqref{equalus} belongs to $\K$.  
 \item  The function $p \mapsto \tilde{u}(x^{\star},p)$ is convex in a neighborhood of $p^{\star}$.
   \end{enumerate}
Moreover, whenever the derivatives exist, they satisfy 
$\nabla_p \tilde{u}=- (\partial_x \tilde{u} )\hat{q}$.
 \end{udiff}

We remark that the equation $\nabla_p \tilde{u}=-(\partial_x \tilde{u})\hat{q}$ holds also for exponential  utilities (see\footnote{Actually,  \cite[Theorem 3.1]{IlJoSir:05} states that $\nabla_p \tilde{u}=\tilde{q}$;
the missing minus sign in front of $\tilde{q}$ (which, in \cite{IlJoSir:05}, is called $\lambda^{\star}$) is a typo, whereas the term $\partial_x \tilde{u}$ is  missing because  in this case $\partial_x \tilde{u}=1$, as it  follows from \cite[Theorem 4.1]{IlJoSir:05}.} \cite[Theorem 3.1]{IlJoSir:05}), and that it can be derived heuristically in a simple fashion.
 Indeed, $p$ is a marginal price at $(x,q)$ iff the agent with endowment $(x,q)$ could achieve no gains by trading in derivatives, so
\be
\label{MargPrMinimize}
p\in \mathcal{P}(x,q) \text{ \, iff \,  } \tilde{u}(x+qp,p)=u(x,q)= \min_{p'} \tilde{u}(x+qp',p'),
\ee
which  allows us to characterize marginal prices using $\tilde{u}$ instead of $u$. It follows from (\ref{MargPrMinimize}) that the function $g(p')=\tilde{u}(x+qp',p')$ has gradient zero at $p'=p$.
Since $p\in \mathcal{P}(x,q)$ implies $q=\hat{q}(x+qp,p)$, we obtain $0=\nabla_{p'} g (p)  =((\partial_x \tilde{u}) \hat{q}+\nabla_p \tilde{u}) (x+qp,p) $ as desired. Notice however how the above line of reasoning does not clarify at which points the equality is satisfied.

\section{Results on existence and uniqueness}
\label{Existence and Uniqueness}

In this section we will state two theorems which subsume Theorem   \ref{thmerge}; the first deals with the existence and uniqueness of the solution of the problem (\ref{prrendtilde}), and the second solves problems  (\ref{prrend}) and \eqref{equalus} and describe their relationship with problem (\ref{prrendtilde}).

We first need to introduce the dual problems.
 If we define
 the set $\mathcal{\bar{L}}$ to be the polar of $-\mathcal{\bar{K}}$:
 \bd
 \label{L}
 \mathcal{\bar{L}}:=-\mathcal{\bar{K}}^o:=\{ v\in\r^{n+1} : vw \geq 0
 \textrm{ for all } w \in \mathcal{\bar{K}} \} , \ed then clearly
 $\mathcal{\bar{L}}$ is a closed convex cone. We will denote by
 $\mathcal{L}$ its relative interior, so  $\mathcal{\bar{L}}$ is the closure of  $\mathcal{L}$.
Given an arbitrary vector $(y, r)\in \r \times \r^{n}$,
 we denote by $\mathcal{Y}(y,r)$ the set of non-negative
 super-martingales $Y\in \mathcal{Y}(y)$ such that the inequality \bd
\textstyle  \mathbb{E}[Y_T (X_T + qf)] \leq xy + qr \ed holds whenever $(x,
 q)\in \mathcal{\bar{K}}$ and $X \in \mathcal{X}(x, q)$; it is easy to show that  
$\mathcal{Y}(y,r)$ is non-empty if and only if $(y,r)\in \mathcal{\bar{L}}$ (see \cite[Remark 5]{Sio12:ArbFreePr}).

 We now define the problems dual\footnote{The duality in \eqref{dualpr} is with respect to the variable $y$  only, with $p$ playing the role of a parameter.} to (\ref{prrendtilde}) and to 
 (\ref{prrend}) to as follows\footnote{We use the convention that the $\sup$ (inf) over an  empty set takes the value $-\infty$ ($+\infty$).}:
 \be
 \label{dualpr}
 \tilde{v}(y,p):=\inf_{Y\in \mathcal{Y}(y,yp)} \mathbb{E}[V(Y_T)] , \quad
 y \in \r 
\ee
and
\be
 \label{dualprrend}
 v(y,r):=\inf_{Y\in \mathcal{Y}(y,r)} \mathbb{E}[V(Y_T)] , \quad
 (y,r)\in \r \times \r^n ,
 \ee
 where  $p\in \P$ is the vector of prices of the contingent claims
 $f$. We notice that trivially $ \tilde{v}(y,p)=v(y,yp)$, and recall the following  relationships between $w,  \tilde{w}, u$ and $v$:  $u(x,0)=w(x)$ (which follows from $\mathcal{X}(x,0)=\mathcal{X}(x)$) and $ \tilde{w}(y)=\min_{p\in \P} v(y,yp)$ (see \cite[Lemma 2]{HugonKram:04}). Moreover, recall that  $\L=\{(y,yp): y>0 \text{ and } p\in \P \}$, or equivalently $\mathcal{P}=\{p:(1,p)\in\mathcal{L}\}$ (see \cite[Lemma 3]{Sio12:ArbFreePr}).
The following  theorem is an
 analogue of the results found in \cite{KramSch:03}. 

 \newtheorem{th2}[thmerge]{Theorem}
 \begin{th2}
   \label{th2}
   Assume that $p$ is an arbitrage-free price for $f$, that conditions
   (\ref{na}), (\ref{inada}) and (\ref{domination}) hold, and
   that 
\begin{align}
\label{dualValFnFinite}
   \tilde{w}(y) <\infty \textrm{ for all } y>0.  \end{align}

   Then one has:
   \begin{enumerate}
   \item \label{biconj1}
The value functions $\tilde{u}$ and $-\tilde{v}$ are finite,
     continuously differentiable, strictly increasing and strictly
     concave on $(0,\infty)$ and satisfy: \bd \tilde{u}'(0):=\lim_{x
       \to 0}\tilde{u}'(x)=\infty , \qquad \tilde{v}'(\infty):=\lim_{y
       \to \infty}\tilde{v}'(y)=0 \, , \ed \bd
     \tilde{u}'(\infty):=\lim_{x \to \infty}\tilde{u}'(x)=0 , \qquad
     \tilde{v}'(0):=\lim_{y \to 0}\tilde{v}'(y)=-\infty \, , \ed as
     well as the bi-conjugacy relationships:
     \begin{align}
       \tilde{u}(x)=\min_{y>0} (\tilde{v}(y)+xy) , \qquad x>0, \nonumber \\
       \tilde{v}(y)=\max_{x>0} (\tilde{u}(x)-xy) , \qquad y>0
       \nonumber
     \end{align}
     where the  optimizers are unique and given by $\tilde{y}=\tilde{u}'(x)$ and
     $x=-\tilde{v}'(\tilde{y})$.
\item The solution $(\tilde{X}(x),\tilde{q}(x))$ to
  (\ref{prrendtilde})   exists for any $x > 0$, and $-\tilde{X}(x)$ is an acceptable wealth process. The final payoff $\tilde{X}_T(x)+\tilde{q}(x)f$ is unique.
   \item  For any $y>0$, $\mathcal{Y}(y,yp) \neq \emptyset$ and the     solution 
$\tilde{Y}(y)$ to (\ref{dualpr}) exists and is unique.
\item \label{Ybdd}
If $\tilde{Y}_T(y)$ is  bounded (or equivalently if $\tilde{X}_T(x)+\tilde{q}(x)f \geq \e>0$, for $x$ given by  $\tilde{y}=\tilde{u}'(x)$), $\tilde{Y}(y)$ equals the density process of an equivalent local-martingale measure and $(x-\tilde{q}(x)p,\tilde{q}(x)) \in \K$.
    \item
     \label{deriv}
     If $x>0$ and $\tilde{y}=\tilde{u}'(x)$ the optimizers of
     (\ref{prrendtilde})    and     (\ref{dualpr}) satisfy
     \begin{align}
       \tilde{Y}_T(\tilde{y})=U'(\tilde{X}_T(x)+\tilde{q}(x)f) \, , \nonumber \\
       \mathbb{E}[\tilde{Y}_T(\tilde{y})(\tilde{X}_T(x)+\tilde{q}(x)f)]=x \tilde{y} \,
       . \nonumber
     \end{align}
   \end{enumerate}
 \end{th2}

To the best of our knowledge, the simple observation that the dual optimizer is a martingale if its terminal value is bounded is new; our proof also applies to the context of  \cite{KramSch:03}, mutatis mutandis.

It trivially follows from \cite[Theorem 3.2]{KramSch:99}, that a convenient sufficient
 condition for the validity of \eqref{dualValFnFinite}  is that 
 the asymptotic elasticity of $U$ is strictly less than one, and that 
$w(x) <\infty \textrm{ for some } x>0. $

 We now solve problems (\ref{prrend}), and (\ref{equalus}),  and elucidate their relationship with problem  (\ref{prrendtilde}).

 \newtheorem{comparison} [thmerge] {Theorem}
 \begin{comparison}
\label{comparison} 
Under the assumptions of Theorem \ref{th2}, the following holds:
\begin{enumerate}
  \item 
\label{solexistsexplicit}
For any $x>0$ the solutions $(x-\hat{q}p,\hat{q}) $ to (\ref{equalus})
     exist, belong to $dom(\partial u)$, and are given by $-\partial    v(\tilde{y},\tilde{y} p)$, where $\tilde{y}=\tilde{u}'(x)$.
\item \label{existsuniquesol}
For every $(x,q) \in \{ u >-\infty \}$, the solution $X(x,q)$ to  (\ref{prrend})  exists and is unique.
\item \label{trivialrelsol}
For any $x>0$ the solutions  $(\tilde{X}(x,p),\tilde{q}(x,p))$   to  (\ref{prrendtilde})  are given by 
\bd
\{  (X(x-qp,q),q): (x-qp,q) \text{ solves (\ref{equalus})}  \}.
\ed
\item
\label{bothsoluniq}
The following conditions are equivalent:
\begin{enumerate}
\item \label{1}
The solution  to (\ref{prrendtilde})  is unique for all $x>0$.
\item \label{2}
The solution  to (\ref{equalus}) is unique for all $x>0$.
\item \label{3}
Condition (\ref{wlog}) holds.
\item \label{4}
The function $v$ is differentiable on $\mathcal{L}$.
\end{enumerate}
If these conditions hold, the solution $(x-\hat{q}p,\hat{q})$ to (\ref{equalus}) is given by  $ -\nabla v( \tilde{y},\tilde{y} p)$, where $\tilde{y}=\tilde{u}'(x)$.
\end{enumerate} 
\end{comparison}
 
Notice that Theorem \ref{comparison} allows to compute $\tilde{q}(x)$  explicitly, as long as one can compute $v$; results on how to approximate  $\tilde{q}(x)$ are contained in \cite{KramSirb:06b}.

Although in principle we could base our proof of Theorem \ref{th2}  on the relationship  between the different optimization  problems stated in Theorem \ref{comparison}, we find it much more economical   to reduce problem \eqref{prrendtilde} to a setting where we can apply the abstract results of  \cite{KramSch:99}, \cite{KramSch:03};
this approach has the additional advantage of automatically providing an alternate version of Theorem \ref{th2} which holds under a different set of hypotheses. In fact, we can rely on \cite[Theorem 3.1]{KramSch:99} to get a weaker version\footnote{The interested reader can easily write down the alternate statement after comparing \cite[Theorem 3.1]{KramSch:99} with  \cite[Theorem 3.2]{KramSch:99}.}  of Theorem \ref{th2} in the case where we do not assume that $\tilde{w}$ is finite, but only that $\tilde{u}(x)<\infty$ for some $x>0$.
 The next remark shows that this assumption is equivalent to the following more natural and at first sight weaker condition (\ref{finiteu}), and thus also to the stronger looking condition that $\tilde{u}$ is finite on $(0,\infty)$.

 \newtheorem{th1}[thmerge]{Remark}
 \begin{th1}
   \label{th1}
   Assume that conditions
   (\ref{na}), (\ref{inada}), (\ref{domination}) hold, that $p\in \P$ and
   \be
   \label{finiteu}   w(x) <\infty \textrm{ for some } x>0. 
  \ee
  Then $\tilde{u}(x)<\infty$ for all $x>0$.
 \end{th1}

 \textsf{ PROOF.}  As explained in \cite[Remark 6]{HugonKram:04}, we can
 assume without loss of generality that (\ref{wlog}) holds.
Notice that $u$ is concave and $u>-\infty$ on
 $\mathcal{K}$ (since $u(x,0)=w(x)$, this follows easily from \eqref{domination}: see \cite[Theorem 1]{HugonKram:04}).
 It follows that $u$ is a proper concave function, and so it is bounded above by an
 affine function. Since \cite[Lemma 3]{Sio12:ArbFreePr}  shows that, for any $x>0$, the set 
 $\{(x-qp,q)\in \mathcal{\bar{K}}: q\in \r^n \}$ is bounded, (\ref{equalus}) implies
 $\tilde{u}(x)<\infty$  for all $x>0$. $\Box$ \newline

 \section{Proofs of existence and uniqueness}
 \label{Optimal investment with stocks and derivatives}

In this section we prove Theorem \ref{th2} by making use of \cite[Theorem 4]{KramSch:03}.
We first need to introduce some notation: define
 \bd \mathcal{C}(x,q):=\{g\in L_+^0 : g\leq
 X_T+qf \textrm{ for some } X \in \mathcal{X}(x,q)\} , \ed and \be
 \label{Ctilde}
 \mathcal{\tilde{C}}(x,p):=\bigcup_{\, q
   \in \r^n }
 \mathcal{C}(x-qp,q) =\bigcup_{\, q
   \in \r^n : (x-qp,q)\in \mathcal{\bar{K}}}
 \mathcal{C}(x-qp,q) .  \ee We will often write  $\mathcal{\tilde{C}}(x)$ (resp.  
$\mathcal{\tilde{C}}$ ) as a
 shorthand for $\mathcal{\tilde{C}}(x,p)$ (resp. $\mathcal{\tilde{C}}(1,p)$),
 and we observe that
 $x\mathcal{\tilde{C}}=\mathcal{\tilde{C}}(x)$.
  
 Define $\mathcal{D}(y,r)$ to be the set of positive random variables
 dominated by the final value of some element of $\mathcal{Y}(y,r)$,
 i.e.,  \be \mathcal{D}(y,r):=\{h\in L_+^0 : h\leq Y_T \textrm{ for
   some } Y \in \mathcal{Y}(y,r) \} \ee We will write
 $\tilde{\mathcal{D}}$ as a shorthand for $\mathcal{D}(1,p)$, and we
 observe that $\mathcal{D}(y,r)\neq \emptyset$ if and only if $(y,r)\in \mathcal{\bar{L}}$ (see \cite[Remark 5]{Sio12:ArbFreePr}), and  $y\mathcal{\tilde{D}}=\mathcal{D}(y,yp)$.

 We recall here two facts proved in \cite[Lemmas 8 and 9]{HugonKram:04}.  Let
 $\mathcal{M}'$ be the set of equivalent local-martingale measures $Q$
 such that the maximal process $X'$ that appears in (\ref{domination})
 is a uniformly integrable martingale under $Q$, and let
 $\mathcal{M}'(p)$ be the subset of measures $Q \in \mathcal{M}'$ such
 that $\mathbb{E}_Q[f]=p$.  If $(1,p)\in \mathcal{L}$ and conditions
 (\ref{na}) and (\ref{domination}) hold, then \be
 \label{M'}
 \mathcal{M}^{'}(p)\neq \emptyset \textrm{,\ and if \,} Q \in
 \mathcal{M}'(p) \textrm{\, then \,} dQ/dP \in \mathcal{D}(1,p).  \ee
 
 We are now ready to prove the analogue of \cite[Proposition 3.1]{KramSch:99}

 \newtheorem{bipolar}[thmerge]{Theorem}
 \begin{bipolar}
   \label{bipolar}
   Assume that $p\in \P$ and that
   conditions (\ref{na}), (\ref{domination}), (\ref{wlog}) hold.
 Then $\mathcal{\tilde{C}}$ is bounded in
   $L^0(\Omega,\mathcal{F},P)$ and it contains the constant function
   $g=1$.  The sets $\mathcal{\tilde{C}}$ and $\mathcal{\tilde{D}}$
   satisfy the bipolar relations: \be
   \label{bipolar1}
   g\in \mathcal{\tilde{C}} \iff g\in L_+^0 \, \textrm{ and } \,
   \mathbb{E}[gh] \leq 1 \quad \forall h \in \mathcal{\tilde{D}} \,
   \ee \be
   \label{bipolar2}
   h\in \mathcal{\tilde{D}} \iff h\in L_+^0 \, \textrm{ and } \,
   \mathbb{E}[gh] \leq 1 \quad \forall g \in \mathcal{\tilde{C}} \, .
   \ee
 \end{bipolar}

 \textsf{ PROOF OF THEOREM \ref{bipolar}. }  The implication
 $\Rightarrow$ in (\ref{bipolar1}) and in (\ref{bipolar2}), and the
 inclusion $1 \in \mathcal{\tilde{C}} $ follow directly from
 definitions.  Now we use (\ref{M'}). If $g\in \mathcal{\tilde{C}}$ and  $Q \in \mathcal{M}'(p)$   then $\mathbb{E}_Q[g] \leq 1$; it
 follows that $\mathcal{\tilde{C}}$ is bounded in $L^1(Q)$, and so
 also in $L^0(P)$, since $Q$ is equivalent to $P$.
  
 To finish the proof of (\ref{bipolar2}) assume that $h$ is a
 non-negative random variable such that $\mathbb{E}[gh] \leq 1 \quad
 \forall g \in \mathcal{\tilde{C}} $. Then, in particular,
 $\mathbb{E}[X_T h] \leq 1 \quad \forall X \in \mathcal{X}(1)$, so
  \cite[Proposition 3.1]{KramSch:99} implies the existence of a process
 $Y \in \mathcal{Y}(1)$ such that $h\leq Y_T$.  Define the process $Z$
 by setting \bd  Z_t:= \left\{
   \begin{array}{ll}
     Y_t & \textrm{ if } t<T \\
     h & \textrm{ if }  t=T .
   \end{array}
 \right.  \ed Then $Z$ belongs to $\mathcal{Y}(1)$ and so, since
 $\mathcal{C}(x,q) \subseteq (x+qp)\mathcal{\tilde{C}}$, it belongs to
 $\mathcal{Y}(1,p)$. This proves $h \in \mathcal{\tilde{D}}$, i.e., the implication
 $\Leftarrow$ in \eqref{bipolar2}.
 
 To conclude, let us prove that $\mathcal{\tilde{C}}$ is closed with
 respect to the convergence in measure; the version of the bipolar
 theorem found in \cite{BranSch:99} and \eqref{bipolar2} then yield (\ref{bipolar1}).  So
 take $g_n \in \mathcal{C}(1-pq_n , q_n)$ and assume without loss of
 generality that $g_n$ converges almost surely to $g$, and let's prove
 that $g \in \mathcal{\tilde{C}}$.  Since \cite[Lemma 3]{Sio12:ArbFreePr} implies
 that $(1-pq_n ,q_n)$ is bounded, passing to a subsequence we can
 assume that $q_n$ is converging to some $q$, so \cite[Lemma 4]{Sio12:ArbFreePr}
 shows that $g\in \mathcal{\tilde{C}}$ $\Box$ \newline

   We will need the following simple remark, which follows from the fact that a wealth process $X$ is maximal iff there is a measure $Q \in \mathcal{M}$ such that  $X$ is a $Q-$uniformly integrable martingale (see \cite[Theorem 2.5]{DelbSch:97}).
 \newtheorem{ind0} [thmerge] {Remark}
 \begin{ind0}
\label{ind0}
If conditions (\ref{na}) and (\ref{domination}) hold, any acceptable wealth process with zero initial value and non-negative terminal wealth is indistinguishable from zero, i.e., $ \mathcal{X}(0,0)=\{0\}$
\end{ind0}

 \textsf{ PROOF OF THEOREM \ref{th2}.  }  As explained in
 \cite[Remark 6]{HugonKram:04}, we can assume without loss of generality that
 (\ref{wlog}) holds.  Since clearly we have \bd
 \label{optmC} \textstyle
 \tilde{u}(x)=\sup_{g \in \,\, \mathcal{\tilde{C}}} \mathbb{E}[U(xg)]
 , \qquad \tilde{v}(y)=\inf_{h\in \mathcal{\tilde{D}}}
 \mathbb{E}[V(yh)] , \ed Theorem \ref{bipolar} puts us in a position
 to apply \cite[Theorem 4]{KramSch:03}, as long as we prove that
 $\tilde{v}$ is finite at all points under the assumption that
 $\tilde{w}$ is; this follows from \cite[Lemma 3]{Sio12:ArbFreePr} and \cite[Lemma 2]{HugonKram:04}.
  
Since problem (\ref{dualpr}) is a particular case of problem (\ref{dualprrend}) (resp. because of item \ref{trivialrelsol} of Theorem \ref{comparison}),
 it is enough to prove item \ref{Ybdd} (resp. the fact minus the optimal wealth process is acceptable)  in the context of optimal investment with random endowment; this will be done in Theorem \ref{duality}, and involves of course no circular argument.

\section{Optimal investment with random endowment}
\label{Optimal investment with random endowment}

In this section we rely on the upper-semi-continuity of $u$ (which was proved in \cite{Sio12:ArbFreePr}) to extend  \cite[Theorem 2]{HugonKram:04}  by considering also the behavior on the boundary of $\mathcal{K}$ and $\mathcal{L}$. 
In particular, we show that the sub-differential of  $v$ is empty on the boundary of $\L$, that $u$ and $v$ are convex conjugates on $\r^n$ and that  if the solution $ Y_T(y,r)$ of the dual problem is a bounded random variable then $Y(y,r)$ equals the density process of an equivalent local-martingale measure and $-\partial v(y,r)\in \K$.

We will denote by $Im(\partial f)$ the image of the sub-differential of
a concave function $f:\r^{n+1} \to (-\infty,\infty]$ (or of a convex
and $[-\infty,\infty)$ valued function), by $dom(\partial f)$ its
domain $\{z: \partial f(z)\neq \emptyset \}$, and by \bd
\textstyle \frac{\partial^+ f}{\partial w}(z):=\lim_{t \to 0+}
\frac{f(z+tw)-f(z)}{t} \ed the right sided directional derivative of
$f$ at $z$ in the direction of $w$.

\newtheorem{duality}[thmerge]{Theorem}
\begin{duality}
  \label{duality}
  Under the assumptions of Theorem \ref{th2} the following holds:
  \begin{enumerate}
  \item
    \label{item1}
    The functions $u$ and $-v$ defined on $\r^{n+1}$ have values in
    $[-\infty,\infty)$, are concave and upper semi-continuous and
    satisfy the bi-conjugacy relationships:
    \begin{align}
      \label{bi-conjugacy1} 
      u(x,r)=\inf_{(y,r)\in \r^{n+1}} \left(v(y,r)+xy+qr\right), \qquad (x,q)\in \r^{n+1},  \\
      \label{bi-conjugacy2}
      v(y,r)=\sup_{(x,q)\in \r^{n+1}} (u(x,r)-xy-qr) , \qquad (y,r)
      \in \r^{n+1} .
    \end{align}
  \item
    \label{dom(partialu)}
    $\mathcal{K}\subseteq dom(\partial u) \subseteq \{u >-\infty \}
    \subseteq \mathcal{\bar{K}}$, and all inclusions can be strict.
  \item
    \label{dom(partialv)}
    $\mathcal{L}= dom(\partial v) \subseteq \{v <\infty \} \subseteq
    \mathcal{\bar{L}}$. In particular $\partial v(\mathcal{L})=
    dom(\partial u)$ and $\mathcal{L}=\partial u(\mathcal{\bar{K}})$
    and if $(y,r)$ belongs to the relative boundary of $\mathcal{L}$
    then $\partial v(y,r)=\emptyset$ and $\frac{\partial^+ v}{\partial
      w}(y,r)=-\infty$ for every $w\in \mathcal{L}-(y,r)$.
  \item
    \label{existopt}
    For all $(x,q) \in \{u>-\infty\}$ there exists a unique maximizer
    $X(x,q)$ of (\ref{prrend}) and for all $(y,r) \in
    \{v<\infty\}$ there exists a unique minimizer $Y(y,r)$ of
    (\ref{dualprrend}). 
\item \label{BddImpliesMart}
The process  $-X(x,q)$ is acceptable, and if    $Y_T(y,r)$ is bounded (or equivalently if $X_T(x,q)+qf \geq \e>0$ for $(x,q)$ such that $(y,r)\in \partial u(x,q)$) then $Y(y,r)$ equals the density process of an equivalent local-martingale measure and $-\partial v(y,r)\in \K$.
  \item
\label{optimizdual}
 If  $(x,q)\in dom(\partial u)$ and $(y,r)\in \partial u(x,q)$, the terminal values of the
    optimizers are related by \be
    \label{optim1}
    Y_T(y,r)=U'(X_T(x,q)+qf) , \ee \be
    \label{optim2}
    \mathbb{E}[Y_T(y,r)(X_T(x,q)+qf)]=xy+qr .  \ee
  \end{enumerate}
 
  \end{duality}
  \textsf{ PROOF. } 
Since $u$ is concave and takes real values on the open set $\mathcal{K}$
(see  \cite[Lemma 2, Theorem 2]{HugonKram:04}), $u$ never takes the value $\infty$ and the inclusions in item \ref{dom(partialu)}  hold.
 Analogously, since $v$ is convex and takes real values on the open set $\mathcal{L}$ (see  \cite[Lemma 2]{HugonKram:04}), $v$ never takes the value $-\infty$, yielding the chain of inclusions in item \ref{dom(partialv)} other than $dom(\partial v) \subseteq \mathcal{L}$, which will be proved later.

By  \cite[Theorem 6]{Sio12:ArbFreePr},  $u$ is upper-semi-continuous, and there exists a unique maximizer of (\ref{prrend}) for any $(x,q) \in \{u>-\infty\}$. 
 To prove that $v$ is lower semi-continuous and that there exists a unique
minimizer for any $(y,r) \in \{v<\infty\}$ let $h_n\in \mathcal{D}(y_n,r_n)$ for some converging sequence
$(y_n,r_n)$, and define $s:=\sup_n y_n <\infty$. Then $h_n \in
\mathcal{D}(s)$, so \cite[Lemma 3.2]{KramSch:99} gives that the
sequence $(V^-(h_n))_{n\geq 1}$ is uniformly integrable. The proof of
 \cite[Theorem 3]{Sio12:ArbFreePr} then applies, mutatis mutandis, completing the proof of item \ref{existopt}.

To  prove  the bi-conjugacy
  relationships (\ref{bi-conjugacy1}) and (\ref{bi-conjugacy2}), call $\bar{u}$ the function defined by the right-hand
  side of (\ref{bi-conjugacy1}). Since $v=\infty$ outside of
   $\mathcal{\bar{L}}$, the infimum defining $\bar{u}$ can equivalently
  be taken over $\mathcal{\bar{L}}$ instead of $\r^{n+1}$, and \cite[Theorem
  7.5]{Rock:70} implies that we can equivalently
  replace $\mathcal{\bar{L}}$ with $\mathcal{L}$ (analogously we can
  replace $\r^{n+1}$ with $\mathcal{\bar{K}}$ or $\mathcal{K}$ in (\ref{bi-conjugacy2})). Then the concave and upper semi-continuous functions 
  $u$ and $\bar{u}$, which are defined on $\r^{n+1}$, never take the
  value $\infty$ and coincide on
  $\mathcal{K}$ (as shown in \cite[Theorem 2]{HugonKram:04}) and so on
  $\mathcal{\bar{K}}$ (by  \cite[Theorem 7.5]{Rock:70})).  By
  definition $u$ is identically $-\infty$ outside $\mathcal{\bar{K}}$;
  let us show that this is also true of $\bar{u}$, so they coincide
  everywhere. If $(x,q) \notin \mathcal{\bar{K}}$ one can find
  $(y,r)\in \mathcal{L}$ such that $xy+qr<0$, and so \bd 
   \frac{\bar{u}(x,q)}{n} \leq  \,  \left( \frac{v(n(y,r))+ n(xy+qr)}{n}
  \right).  \ed The thesis then follows taking limits for $n \to \infty$, using l'Hospital rule and the
  fact that, by Theorem \ref{th2}, the function $\tilde{v}(\lambda)=v(\lambda(1,r/y))$  satisfies $\lim_{\lambda \to \infty} \tilde{v}'(\lambda)=0$ ($r/y$ is an arbitrage-free price: see \cite[Lemma 3]{Sio12:ArbFreePr}).  This concludes the  proof of (\ref{bi-conjugacy1}), and now (\ref{bi-conjugacy2}) follows  from the fact that $v$ is convex, proper and lower semi-continuous, concluding the proof of item \ref{item1}.

Let us prove item \ref{optimizdual}. If
  $(y,r)\in \partial u(x,q)$ then $(x,q)\in \partial v(y,r)$ and so, by
  definition of sub-differential, $(x,q) \in \{u>-\infty\}$ and $(y,r) \in
  \{v<\infty\}$. Item \ref{existopt} implies the existence of the
  optimizers $X(x,q)$ and $Y(y,r)$, and we have \be
  \mathbb{E}[\, \big|V(Y_T(y,r))+X_T(x,q)Y_T(y,r)-U(X_T(x,q))\big| \, ]=
  \nonumber \ee \be
  \mathbb{E}[V(Y_T(y,r))+X_T(x,q)Y_T(y,r)-U(X_T(x,q))]
  \nonumber \ee \be \leq v(y,r)+xy+qr-u(x,q)=0 , \nonumber \ee which
  implies (\ref{optim1}) and (\ref{optim2}). 

Let us now prove item \eqref{BddImpliesMart}.
Observe that
 there exists  $X\in \mathcal{X}(x,q)$ such
 that $-X$ is acceptable and $X_T\geq X_T(x,q)$ (see \cite[Lemma 2]{Sio12:ArbFreePr}); since $X(x,q)$ is an optimizer, this implies
 $X_T=X_T(x,q)$. Then $X(x,q)-X \in
 \mathcal{X}(0,0)$, so Remark \ref{ind0} gives that $-X(x,q)$ is acceptable.
Now, fix  $(y,r)\in \partial u(x,q)$ and assume that $Y_T(y,r)\leq c<\infty$, which  (\ref{optim1}) implies to be equivalent to  $X_T(x,q)+qf\geq \varepsilon>0$ for  $\varepsilon:=(U')^{-1}(c)$; then
\begin{align*} \textstyle
X(x,q)-\varepsilon=:X \in \mathcal{X}(x-\varepsilon,q), \text{ and in particular } (x-\varepsilon,q)\in \Kbar .
\end{align*}

Since  $(1,0)\in \K$ (see  \cite[Lemma 6]{HugonKram:04}) and $\Kbar$ is a convex cone,  $(x,q)=(x-\varepsilon,q)+\varepsilon(1,0)\in \Kbar +\K \subseteq \K$, i.e., $-\partial v(y,r)\subseteq \K$.
From the definition of $\mathcal{Y}(y,r)$ and $X \in \mathcal{X}(x-\varepsilon,q)$ it follows that 
\bd \textstyle
\mathbb{E}[Y_T(y,r)(X_T+qf)]\leq (x- \varepsilon)y+qr , 
\ed
 so (\ref{optim2}) and the definition of $X$ imply that  $\mathbb{E}[-\varepsilon Y_T(y,r)]\leq  -\varepsilon y .$ Thus $\mathbb{E}[Y_T(y,r)]\geq y $, the super-martingale $Y(y,r)$ is actually a martingale, and so $Q$ defined by $dQ=Y_T(y,r)dP$ is a probability measure whose density process equals $Y(y,r)\in \mathcal{Y}(y)$; since $S$ is locally-bounded, $Q$ is an equivalent local-martingale measure.

Let us show with examples that the inclusions in item \ref{dom(partialu)}  
can be strict; take $U$ to be a
  power utility of exponent $\alpha$ and considering a market with no
  stocks and one contingent claim uniformly distributed in
  $[-1,1]$. Then one can easily compute explicitly $u(x,q)$ for all $\alpha$'s, and show the following. When $\alpha$ is smaller than $-1$, $dom(u)=\K$, so the last inclusion can be strict. When $\alpha$ is between $-1$ and $0$,   $dom(u)=\Kbar \setminus \{(0,0)\}$ and $ dom(\partial u)=\K$, so the second inclusion can be strict. Finally, when $\alpha$ is between $0$ and $1$, $ dom(\partial u)=\Kbar \setminus \{(0,0)\}$, so the first inclusion can be strict.

To conclude the proof of item \ref{dom(partialv)}, we only need to prove that $ \partial u(\mathcal{\bar{K}})\subseteq   \mathcal{L}$, since item
  \ref{item1} implies that $\partial u(\r^n)=  dom(\partial v)$ and $\partial v(\r^n)=  dom(\partial u)$   (see \cite[Theorem 23.5]{Rock:70}), and the result on the partial derivatives of $v$ follows then from   \cite[Theorem 23.3]{Rock:70}. 
Notice first that 
\bd
\partial u(\mathcal{\bar{K}})=  dom(\partial v)\subseteq
  \mathcal{\bar{L}}.
\ed 
Now $ \partial u(\mathcal{\bar{K}})\subseteq   \mathcal{L}$ follows exactly as in the proof of    \cite[Theorem 2]{HugonKram:04} once we notice that the following relationship holds for every  $(x,q) \in \mathcal{\bar{K}}$ (and not just for $(x,q) \in \mathcal{K}$): for any non-negative measurable function  $g$,
 \be
\label{dualityCxq}
 g \in \mathcal{C}(x,q) \iff \mathbb{E}[gh] \leq
  1 \quad \forall h \in \mathcal{\tilde{D}}(x,q), \ee
where by definition
 \bd   A(x,q):=\{ (y,r)\in \mathcal{\bar{L}} : xy+qr \leq 1 \} , \ed \bd \textstyle
  \mathcal{\tilde{D}}(x,q):=\bigcup_{(y,r)\in A(x,q) }
  \mathcal{D}(y,r). \ed 
 Indeed, equivalence (\ref{dualityCxq}) corresponds to   \cite[Proposition 1, Eq. (26)]{HugonKram:04} which, although stated only for $(x,q) \in \mathcal{K}$,  holds
  (with the same proof)  for every  $(x,q) \in \mathcal{\bar{K}}$ (this is true also for \cite[Lemma 10]{HugonKram:04},  on which   \cite[Proposition 1]{HugonKram:04} relies).
   $\Box$ \newline

Trivially there is a condition which is equivalent to the domain of $u$ being the whole of $\Kbar$: that is, $U(0):=U(0+)$ needs to be real valued.
 Indeed 
\be
\label{trivialineq}
u(x,q)\geq u(0,0) \textrm{\, at all \, } (x,q)\in \Kbar ,
\ee
 so $dom(u)=\Kbar$ iff $u(0,0)\in \r$, and Remark \ref{ind0} implies that $u(0,0)=U(0)$.
 
\section{Relation between the optimization problems}
\label{Relation between the two optimization problems}
In this section we prove Theorem \ref{comparison}.

\textsf{ PROOF OF THEOREM \ref{comparison}. }
Item \ref{existsuniquesol} is part of item \ref{existopt} in Theorem \ref{duality}, and item \ref{trivialrelsol} is trivial.
 For the proof of item \ref{solexistsexplicit}, fix $x$ and $p$, and let $f$ be the function $f(q):= - u(x-pq,q)$, which is proper convex and lower semi-continuous (by Theorem \ref{duality}); its minimizer is then given by $\partial f^*(0)$ (see \cite[Formula (1.4.6), Chapter E]{HiriarLemar:01}), where $f^*$ is the Fenchel-Legendre transform of $f$. To compute $f^*$ we write $f=g \circ A$ with $g(a,b):= -u (- (a,b))$, $A(q):=(-x+pq,-q)$, and apply \cite[Chapter E, Theorem 2.2.3]{HiriarLemar:01} to find 
\bd
f^*(\mu)=min_{(y,r)} \{ v(y,r)+xy : yp-r= \mu \}.
\ed
The previous expression for $f^*$ allows us to compute $\partial f^*(0)$ using   \cite[Chapter D, Theorem 4.5.1]{HiriarLemar:01}\footnote{Although stated only for finite convex functions, the theorem holds (with the same proof) for proper convex functions.}, and to find 
\be
\label{minimiz}
\partial f^*(0)=\{q : (x-pq,q) \in -\partial v(y,r) \},
\ee
where $(y,r)$ is any solution of 
\be
\label{constrdualmin} 
min_{(y,r)} \{ v(y,r)+xy : yp-r=0 \}.
\ee
 Item \ref{biconj1} of Theorem \ref{th2} then implies that
problem (\ref{constrdualmin}) has as unique solution, namely 
$(\tilde{y},\tilde{y} p)$, with  $\tilde{y}=\tilde{u}'(x)$. Stitching the pieces together, we obtain that the function $q\mapsto u(x-pq,q)$ is maximized at the points $(x-p\hat{q},\hat{q})$ in $ -\partial v(\tilde{u}'(x),\tilde{u}'(x) p)\subseteq -\partial v(\r^n)=dom(\partial u)$,
 which concludes the proof of item \ref{solexistsexplicit}.

The equivalence of items \ref{3} and \ref{4} is part of \cite[Lemma 3]{HugonKram:04}, and the  the equivalence of items \ref{1} and \ref{2} follows from items \ref{existsuniquesol} and  \ref{trivialrelsol}.
If item \ref{4} holds, item \ref{solexistsexplicit} shows that  $-\nabla v(\tilde{y},\tilde{y} p)$ is the unique solution to (\ref{equalus}); in particular item \ref{2} holds. 
 To show that item \ref{2} implies item \ref{4} we have to proceed differently\footnote{Since item \ref{2} implies that  $-\nabla v(\tilde{y},\tilde{y} p)$ exists at all $\tilde{y}>0$, but only for the one fixed $p$ which we used in problem (\ref{equalus}), and not for all $p$ such that $(1,p)\in \mathcal{L}$.}; so, let us assume that item  \ref{3} does not hold, and let  $q'\neq 0$ be such that $q'f$ is replicated by a process $X' \in \mathcal{X}(x')$. This implies that $x'=q'p$,  since $p$ is an arbitrage-free price. Now, let $(\tilde{X},\tilde{q})$ be a solution to (\ref{prrendtilde}); 
 then,  $(\tilde{X} +x' - X',\tilde{q}+ q')$ is also a solution to
 (\ref{prrendtilde}) (since it has the same final payoff), so item \ref{1} does not hold. This concludes the proof of item \ref{bothsoluniq}.
 $\Box$ \newline

\section{Continuity}
\label{Continuity}

In this section we state and prove a result that subsumes the part of  Theorem \ref{continuity1} which deals with the continuous dependence on $(x,p)$.

 \newtheorem{continuity3}[thmerge]{Theorem}
 \begin{continuity3}
   \label{continuity3}
  Under the assumptions of Theorem \ref{th2} the maps
   \begin{eqnarray}
     (0,\infty) \times \mathcal{P} & \longrightarrow &  L^0(P) \times
     L^1(P) \times \r  \times \r\nonumber\\
     (x,p) & \mapsto & \left(\tilde{X}_T+\tilde{q}f,
       U(\tilde{X}_T+\tilde{q}f),  \tilde{u}(x,p), \frac{\partial\tilde{u}(x,p) }{\partial x}
     \right)\nonumber 
   \end{eqnarray}

   and
   \begin{eqnarray}
     (0,\infty) \times \mathcal{P} & \longrightarrow &  L^0(P) \times
     L^1(P) \times \r  \times \r \nonumber\\
     (y,p) & \mapsto & \left(\tilde{Y}_T, V(\tilde{Y}_T) , \tilde{v}(y,p), \frac{\partial\tilde{v}(y,p) }{\partial y}
     \right) \nonumber 
   \end{eqnarray}
   are continuous, and if we additionally assume (\ref{wlog}) then the
   map
   \begin{eqnarray}
     (0,\infty) \times \mathcal{P} & \longrightarrow & \r^n \times L^0(P) \nonumber\\
     (x,p) & \mapsto & (\tilde{q},\tilde{X}_T) \nonumber 
   \end{eqnarray}
   is (well defined and) continuous.
 \end{continuity3}
 \textsf{ PROOF: Step 1} 
To obtain the proof for $(\tilde{Y}_T,
 V(\tilde{Y}_T) )$ apply the changes which we will now describe to the
 proof of \cite[Lemma 3.6]{KramSch:99}. Replace $y$ and $y_n$ with
 $(y,yp)$ and $(y_n,y_n p_n)$, $\hat{h}$ with $\tilde{Y}_T$ and the
 function that is there called $v$ (which in this article we denote by
 $\tilde{w}$) with the function $v$ defined in (\ref{dualprrend}).
 Apply
  \cite[Lemma 2]{HugonKram:04} to obtain the finiteness of $v$ from our assumption that $\tilde{w}$ is finite, and
 then use \cite[Lemma 4]{Sio12:ArbFreePr} to obtain $g\in
 \mathcal{D}(y,yp)$. Finally use $\mathcal{D}(y,yp) \subseteq
 \mathcal{D}(y)$ and the fact that $v$, being concave and finite on
 the open set $\mathcal{L} \ni (y,yp)$, is there continuous; this
 concludes the proof that  $(\tilde{Y}_T,
 V(\tilde{Y}_T) )$ is continuous.  The continuity of $\tilde{v}$ is now trivial, and the continuity of its derivative in $y$ follows from
 \cite[Theorem 25.7]{Rock:70}.

 \textsf{ Step 2 } Here we prove the continuity  of $\tilde{u}$.
 Since $\tilde{u}(\cdot,p)$ is concave, it is
 locally Lipschitz, so $\tilde{u}(\cdot,\cdot)$ is continuous if
 $\tilde{u}(x,\cdot)$ is continuous. Since, by Theorem \ref{th2}, $\tilde{u}(x,\cdot)$ is the
 infimum over $y>0$ of the continuous functions $p\mapsto v(y,yp)+xy$,
 it is upper-semi-continuous. To prove its lower semicontinuity define
 the open set \bd V_p:=\{q \in \r^n :(x-qp,q) \in \mathcal{K} \} , \ed
 and observe that, since $u$ is upper-semi-continuous (see  \cite[Theorem 6]{Sio12:ArbFreePr}) and concave,  \cite[Theorem 7.5]{Rock:70} yields \bd u(a,b)=\lim_{\varepsilon \to    0+} u((a,b)(1-\varepsilon)+(1,0)\varepsilon ) \text{ for all } (a,b) \in \mathcal{\bar{K}} , \ed which implies that for any $c \in
 \r$ \be
 \label{u(p)sc}
 \tilde{u}(x,p)\leq c \text{\quad if and only if \quad } u(x-qp,q)\leq
 c \text{ for all } q \in V_p .  \ee Since $u$, being concave, is continuous when
 restricted to the interior $\mathcal{K}$ of its effective domain $\{u\in \r\}$, if $p_n \to p \in \mathcal{P}$ and
 $\tilde{u}(x,p_n)\leq c$ then for all $q \in V_p$ \bd
 u(x-qp,q)=\lim_n u(x-qp_n,q) \leq c , \ed
 so (\ref{u(p)sc}) implies
 $\tilde{u}(x,p) \leq c$, so $\{\tilde{u}(x, \cdot) \leq c \}$ is
 closed, i.e., $\tilde{u}(x,\cdot)$ is lower-continuous and so
 $\tilde{u}(\cdot,\cdot)$ is continuous\footnote{We can also give a more elegant proof that $\tilde{u}(x,\cdot)$ is continuous, relying on a hard-to-prove theorem: since $\tilde{v}$ is continuous and $\tilde{u}(x)=\tilde{v}(\tilde{y})+x\tilde{y}$, where $ \tilde{y}= \partial_x \tilde{u}$, the continuity of $\tilde{u}$ follows from the one of $\tilde{y}$. To prove the latter, observe that the continuous bijection $g$ of $(0,\infty)\times \P$ in itself given by $(y,p)\mapsto (-\partial_y \tilde{v}(y,p),p)$ has inverse $g^{-1}(x,p)=(\tilde{y},p)$, and the map $g$ is open, by Brouwer's invariance domain theorem, so $g^{-1}$ is continuous.}.
 The continuity of $\partial_x \tilde{u}$ now follows from \cite[Theorem 25.7]{Rock:70}.
As proved in Theorem \ref{comparison},  
 $\tilde{q}$ is uniquely defined iff (\ref{wlog})  holds, and in this case $\tilde{q} =-\nabla v(\partial_x \tilde{u}, (\partial_x \tilde{u}) \, p)$, so $\tilde{q}$ is  continuous.

 \textsf{ Step 3 } Now we describe the changes one needs to apply to
 the proof of \cite[Lemma 3.6]{KramSch:99} in order to obtain continuity
 of $(\tilde{X}_T+\tilde{q}f, U(\tilde{X}_T+\tilde{q}f))$.  Replace
 $y$ and $y_n$ with $(x,p)$ and $(x_n,p_n)$, $\hat{h}$ with
 $\tilde{X}_T+\tilde{q}f$, $V$ with $-U$ and the function that is
 there called $v$ with the function $\tilde{u}$.  Observe that we can
 assume without loss of generality that (\ref{wlog}) holds, so if
 $\tilde{q}_n:=\tilde{q}(x_n,p_n)$ and $\tilde{q}:=\tilde{q}(x,p)$
 then $\tilde{q}_n \to \tilde{q}$.  Define \bd
 h_n:=\frac{(\tilde{X}_T(x_n,p_n)+\tilde{X}_T(x,p))+(\tilde{q}(x_n,p_n)+\tilde{q}(x,p))f}{2}
 , \ed so that $ h_n \in \mathcal{C}(a_n,b_n)$, where \bd
 (a_n,b_n):=\Big(\frac{x_n+x}{2}-\frac{(\tilde{q}_n
   p_n+\tilde{q}p)}{2}, \frac{\tilde{q}_n+\tilde{q}}{2}\Big) .  \ed
 Since $(a_n,b_n)$ is a convergent sequence, (\ref{domination})
 provides an $\bar{x}>0$ such that $ \mathcal{C}(a_n,b_n)$ is
 contained in $\mathcal{C}(\bar{x},0)$, which is bounded in
 $L^0(P)$. This yields \cite[Formula (3.13)]{KramSch:99} in our case,
 and also allows to apply Kolmos' lemma to construct a sequence
 $(g_n)_{n\geq 1}$ of forward convex combinations of $(h_n)_{n\geq 1}$
 that is converging almost surely to some random variable $g$, which
  \cite[Lemma 4]{Sio12:ArbFreePr} shows to be in $\mathcal{\tilde{C}}(x,p)$.
 Using the continuity of $\tilde{u}$, since $(g_n)_{n\geq 1} \subseteq \mathcal{C}(\bar{x},0)$ we can
 apply \cite[Lemma 1]{KramSch:03} to prove the uniform integrability of
 $U^+(g_n)$. $\Box$ \newline

\section{Asymptotic results and convexity of {\it m}}
\label{asymptotic}

In this section we will conclude the proof of Theorem \ref{continuity1},
addressing the convexity of $m$ and the asymptotics for $\mathcal{P} \ni p_n \to p \notin \P$.
 \newtheorem{Pbdd}[thmerge]{Lemma}
 \begin{Pbdd}
\label{Pbdd}
   If (\ref{na}) and (\ref{domination}) hold, $\mathcal{P}$ is bounded and  convex, and it is open iff (\ref{wlog}) holds.
 \end{Pbdd}
 \textsf{ PROOF.}   The identity $\mathcal{P}=\{p:(1,p)\in\mathcal{L}\}$ (proved in  \cite[Lemma 3]{Sio12:ArbFreePr}) shows  that $\mathcal{P}$ is convex, and is open if and only if condition (\ref{wlog}) is  satisfied (see \cite[Lemma 3]{HugonKram:04}). 
Moreover, assume by contradiction that $\P$ is not bounded, i.e., there exists $p_n \in
 \mathcal{P}$ such that $|p_n|$ is converging to infinity as $n\to
 \infty$, and define \bd q_n:=-p_n/|p_n|^{3/2}. \text{\quad Then }
 1+q_n p_n=1-|p_n|^{1/2} \ed is converging to $ -\infty$, and so it is
 negative for big enough $n$. Since $-\mathcal{\bar{K}}^o=\mathcal{\bar{L}} \ni (1,p_n)$, by definition of polar $(1,q_n) \notin \mathcal{\bar{K}}$, which by
 \cite[Lemma 1]{HugonKram:04} contradicts (\ref{domination}), since $q_n \to
 0$.  $\Box$ \newline

To  obtain our results on the function $m$, we will first study the following auxiliary function $d:\mathcal{L} \longrightarrow
 [0,\infty]$ defined as \be
 \label{d}
 d(w):=\sup\{ |v| : v \in \mathcal{\bar{K}} \text{ and } vw \leq 1\}
 . \ee

 \newtheorem{qsbdd}[thmerge]{Lemma}
 \begin{qsbdd}
   \label{qsbdd}
   Assume conditions (\ref{na}), (\ref{domination}), (\ref{wlog}),
   then
   \begin{enumerate}
   \item \label{dhom}
 $x d(w)=d(\frac{w}{x})$ for every $x>0, w\in
     \mathcal{L}$.
   \item $d$ is finite and the supremum in $(\ref{d})$ is attained.
   \item \label{dsubadd}
$d(w_1+w_2)\leq \min( d(w_1), d(w_2) )$ for all $ w_1,w_2 \in \mathcal{L}$.
   \item $d$ is locally bounded.   \end{enumerate}
 \end{qsbdd}
 \textsf{ PROOF.}  Item one is obvious, and then \cite[Lemma 3]{Sio12:ArbFreePr} 
 implies that $d$ is finite at
 every point, so clearly by compactness the supremum defining $d$ is
 attained. Item three follows from
 $\mathcal{\bar{L}}=-\mathcal{\bar{K}}^o$. 
Finally, let $k=2^{n+1}$ and take  $(h_i)_{i=1}^k \subset \mathcal{L}$ to be the vertices of a cube  $Q$ containing $w$ in its interior, and $\lambda=(\lambda_i)_{i=1}^k$ be convex weights. Let  $h_{\lambda}:=\sum_{i=1}^k \lambda_i
 h_i$  be the generic point in $Q$, and choose $j$ such that $\lambda_j\geq 1/k$; using item  \ref{dhom} and \ref{dsubadd} we obtain \bd d(h_{\lambda})\leq d(\lambda_j h_j)
 \leq k d(h_j) \leq k\max_{1\leq i \leq k} d(h_i)=:C_Q<\infty . \ed
 Thus $d$ is bounded by $C_Q$ on $Q$, so $d$ is locally bounded. 
 $\Box$ \newline
 
 \textsf{ PROOF OF THEOREM \ref{continuity1}.} 
The continuity properties follow from Theorem \ref{continuity3}, and the properties of $\P$ from Lemma \ref{Pbdd}; let us prove the asymptotic properties, assuming $U(\infty)=\infty$. Since $\tilde{u}(\cdot,p)$ is increasing we can assume without loss of
 generality that $x_n=x$. Since $p$ is not an arbitrage-free price, by
 definition there exists an $X \in \mathcal{X}(-qp,q)$ such that the
 inequality $X_T \geq 0$ holds and is strict with strictly positive
 probability. By monotone convergence \bd \lim_n \mathbb{E}[U(x+nX_T)]
 = P(X_T=0)U(x)+ P(X_T>0)U(\infty)=\infty , \ed and so
 $\tilde{u}(x,p)=\infty$ for any $x>0$, and so given an arbitrary $M>0$
 we can find a $q \in \r^n$ such that $u(\frac{x}{2}-qp,q)\geq M$.
 But then, if $\mathcal{P} \ni p_n \to p \notin \P$,
 for $n$ big enough $q(p_n-p)<\frac{x}{2}$ and so \bd
 \tilde{u}(x,p_n) \geq \tilde{u}(\frac{x}{2}+q(p_n-p),p_n) \geq
 u(\frac{x}{2}+q(p_n-p)-qp_n,q)=u(\frac{x}{2}-qp,q) \geq M , \ed which
 proves $\lim_n \tilde{u}(x,p_n)=\infty$.

 If the sequence $q_n:=\tilde{q} (x_n,p_n)$ was converging to some $q
 \in \r^n$, the upper-semicontinuity of $u$ would imply \bd \infty
 =\lim_n \tilde{u}(x,p_n)=\lim_n u(x-q_np_n,q_n) \leq u(x-qp,q) , \ed
 which is not possible since the concave function $u$ is real valued
 on some open set. It follows analogously that $q_n$ can not have any
 convergent subsequence, which implies $\lim_n |\tilde{q}(x_n,p_n)|
 =\infty $; since $|\tilde{q}| \leq m $,  also $m(x_n,p_n)$ diverges.
Let us now prove the results about the function $m$. The positive homogeneity  in $x$ is trivial and, since $\mathcal{P}=\{p:(1,p)\in\mathcal{L}\}$ (see \cite[Lemma 3]{Sio12:ArbFreePr}),  $m(1,p)$  is bounded 
above  by $d(w)$ for $w:=(1,p)$. Thus $m$ is finite, and by compactness this implies that the  supremum in (\ref{m}) is attained. 
To conclude, we only need to show that $m(1,\cdot)$ is convex, as this will also imply the continuity of $m(\cdot,\cdot)$.
Note that 
 \be
 \label{mprime} \textstyle
 m(x,p)=\max \left\{ \, |q| :  (z,q) \in (\r \times \r^n)\cap \mathcal{\bar{K}} \text{ and }  \, (z,q)(1,p)\leq x  \right\},
 \ee
since trivially any maximizer $(\bar{z},\bar{q})$ of \eqref{mprime} satisfies $(\bar{z},\bar{q})(1,p)= x$.
 From Lemma \ref{qsbdd} it follows that, for fixed $w\in \L$, there exists $\e >0 $ s.t.
\be
\label{f(e)<d}
2 \e  \sup \left\{ d(w') : w'\in B_\e (w)\right\}  <1,
\ee
 where $ B_\e (w)$ is the ball of radius $\e$ centered at $w$.
Let $p\in \P$,  $p_0, p_1 \in B_\e (p)$, and define  $w:=(1,p)$ and 
  $w_i :=(1,p_i)$ for $i=0,1$.  Fix a  generic $\l \in (0,1)$, let $$  p_\l := \l p_1 + (1-\l) p_0 \text{ and } w_\l := \l w_1 + (1-\l) w_0=(1,p_\l),$$ and  use (\ref{mprime}) to choose a $v_\l=(z_\l,q_\l) \in (\r \times \r^n) \cap \mathcal{\bar{K}}$  that satisfies $|q_\l|=m(1,p_\l)$ and  $v_\l w_\l \leq 1$, so that necessarily $v_\l w_\l = 1$.
 Suppose that we can build $t_0,t_1 \in \r$ such that
\be
\label{t's}
t_i>0  ,\quad t_i v_\l w_i \leq 1 \text{ for } i=0,1 \text{\, and \, } 1= \l t_1 + (1-\l) t_0 .
\ee
Then, if we define $v_i:=(z_i,q_i):=t_i(z_\l,q_\l)=t_i v_\l$ for $i=0,1$, we have that  $v_i \in \mathcal{\bar{K}}$,  $v_i w_i\leq 1$ and  $v_\l = \l v_1 + (1-\l) v_0$, and so (\ref{mprime}) implies
\bd
m(1,p_\l) = |q_\l| \leq  \l |q_1| + (1-\l) |q_0| \leq  \l m(1,p_1) + (1-\l) m(1,p_0) ,
\ed
which shows that $m(1,\cdot )$ is locally convex and thus\footnote{It is enough to show this in dimension one, where a function is convex iff it is the integral of an increasing function; since a locally increasing functions is increasing, the thesis follows.} convex; so, to conclude we just need to build $t_0,t_1$ that satisfy (\ref{t's}). 
To do so, define $t_1$ such that $t_1 v_\l w_1 = 1$, i.e., 
\bd
t_1:=\frac{1}{1+(1-\l)v_\l (w_1-w_0)}.
\ed
Note that $t_1$ is (well defined and) strictly positive, since  (\ref{f(e)<d}) implies that
\be
\label{|z|<1}
|v_\l (w_1-w_0)| \leq 2 |v_\l | |(w_1-w)|\leq 2 d(1,w_\l) \e <1 .
\ee
Now, define $t_0$ such that $1= \l t_1 + (1-\l) t_0 $ holds. It is easy to show that   (\ref{|z|<1}) implies $t_0>0$, and somewhat lengthy but straightforward\footnote{Just remember to use the identities $w_0=w_\l-\l(w_1-w_0)$ and $v_\l w_\l = 1$.} algebraic manipulations show  that $t_0 v_\l w_0 \leq 1$ is equivalent to $(\l-1) (v_\l (w_1-w_0))^2 \leq 0$, and so it is satisfied.
 $\Box$
 \newline

\section{The one dimensional case}
\label{1dc}
 In this section we prove Theorem \ref{1d}; we will need the following lemma. 

 \newtheorem{1dlemma}[thmerge]{Lemma}
 \begin{1dlemma}
   \label{1dlemma}
      Under the assumptions of Theorem \ref{thmerge}, if there is only one European contingent claim $f$, and
 $ p_1,p_2 \in \mathcal{P}, x>0$, then
   \begin{enumerate}
   \item
     \label{decreas1} If $ p_2< p_1 $ and $\tilde{q}(x,p_1)> 0$
     then $\tilde{q}(x,p_2)> 0$ and $\tilde{u}(x,p_1)<
     \tilde{u}(x,p_2)$.
   \item
     \label{decreas2}
     If $p_2> p_1$ and $\tilde{q}(x,p_1)< 0$ then
     $\tilde{q}(x,p_2)<0$ and $\tilde{u}(x,p_1)<
     \tilde{u}(x,p_2)$.
   \end{enumerate}
 \end{1dlemma}
 \textsf{ PROOF }
 Assume that  $ p_2< p_1 $ and $q_1:=\tilde{q}(x,p_1)> 0$, and let $a_1:=x-q_1p_1$, so that $ u(a_1,q_1) = \tilde{u}(x,p_1)\geq w(x)>-\infty$ and $(a_1,q_1)\in \Kbar$.  In particular $$0<(a_1,q_1)(1,p_2)=x-q_1(p_1-p_2)<x,$$ and so there exists $t_1>1$ s.t. $t_1(a_1,q_1)(1,p_2)=x$.
From  \cite[Lemma 7]{Sio12:ArbFreePr} it follows trivially  that  $t\mapsto u(tx,tq)$ is strictly increasing on $[1,\infty)$ if $(x,q)\in \{u>\infty\}$ and $(x,q)\neq 0$; thus  
\be \textstyle
\label{pineq}
 \tilde{u}(x,p_1)=u(a_1,q_1)<u(t_1(a_1,q_1))\leq  \tilde{u}(x,p_2).
\ee
 Moreover if $q\leq 0$ then $$(x-qp_2,q)(1,p_1)=x+q(p_1-p_2)\leq x$$ and so $u(x-qp_2,q)\leq \tilde{u}(x,p_1)$; it follows that (\ref{pineq}) implies that $q\neq \tilde{q}(x,p_2)$, concluding the proof of item one. Item two follows analogously.  $\Box$ \newline

 \textsf{ PROOF OF THEOREM \ref{1d} } 
As shown in   \cite[Corollary 8]{Sio12:ArbFreePr},  $\mathcal{P}(x,0)$ is the image of the set $\partial u(x,0)$  through the `perspective function' $(y,r)\mapsto r/y$. Since $\partial u(x,0)$  is compact (see \cite[Proposition 4.4.2]{Bert:03}) and convex, this implies that $\mathcal{P}(x,0)$  is also compact and convex\footnote{Because the perspective function is continuous, and sends convex sets to convex sets (as proved in \cite[Section 2.3.3]{BoydVan:04}). Alternatively,  
convexity can also easily be proved directly from the definition of $\P(x,q)$}
 set; thus, there exist $a\leq b$ such that
 $[a,b]=\mathcal{P}(x,0)$. The inclusion $\mathcal{P}(x,0)\subset
 \mathcal{P}$ is proved in  \cite[Theorem 1]{Sio12:ArbFreePr}; from Lemma \ref{Pbdd} it follows that there exist $\underline{p},\overline{p}$ such that 
$\mathcal{P}=(\underline{p},\overline{p})$. 
The continuity of $\tilde{q}$ and $\tilde{u}$ and the asymptotic results follow from Theorem \ref{continuity1}.
 By  definition $p\in \mathcal{P}(x,0)$ iff 
$\tilde{q}(x,p)=0$, and so Lemma \ref{1dlemma} implies that
 $\tilde{q}(x,\cdot)$ is strictly positive on $(\underline{p},a)$, it
 equals zero on $[a,b]$, and it is strictly negative on
 $(b,\overline{p})$. Also, by definition 
$ \tilde{u}(x,p)\geq u(x,0)$ with equality holding iff $p\in \mathcal{P}(x,0)$;  so $ \tilde{u}$ is constant on $ \mathcal{P}(x,0)$, and Lemma \ref{1dlemma} implies that
it is strictly increasing on $(\underline{p},a)$ and is strictly decreasing on
 $(b,\overline{p})$.
 $\Box$ \newline

\section{An example of a non-convex dependence on prices}
\label{non-convex}

In the case of an exponential utility $U$, the function $p \mapsto
 \tilde{u}(x,p)$ is strictly convex (see \cite[Theorem 3.1]{IlJoSir:05}).
 However, in this section we will  show that, in our general framework, convexity does not hold. 

To do so, we will first build a counter-example using a `utility function' $U$ which is not differentiable at $x=1,3$, is convex but not strictly convex, and which satisfies otherwise all the properties we assume in Section \ref{The model} for a utility function. To obtain an example starting with a (differentiable, strictly convex) utility function, it is then enough to consider a sequence of (differentiable, strictly convex)  utility functions $U_n$ converging to $U$ \emph{uniformly}, and  to note that trivially the corresponding maximal expected utilities 
$ \tilde{u}_n$ converge uniformly to $ \tilde{u}$; and so, since 
$ \tilde{u}(x,\cdot)$ is not convex, for some $n$ also $\tilde{u}_n(x,\cdot)$ must be not convex.

In our counter-example we will consider a function $U$ which is affine in $[1/2,1]$, $[1,3]$ and $[3,4]$, and which satisfies 
\bd
U(1)=0, \quad U'(1-)=1000, \quad  U'(1+)=1, \quad  U'(3+)=1/1000,
\ed
so that in particular $U(3)=2.$  We then extend $U$ to $(0,\infty)$ in a way that $U$ is strictly increasing, convex, differentiable at all points other than $x=1,3$, and satisfies Inada conditions (\ref{inada}). We consider a market with no stocks, and one derivative $f$ with distribution given by $P(f=1)=2/3$ and $P(f=-1)=1/3$; the interval of its arbitrage-free prices is clearly $\P=(-1,1)$. Then, the expected utility  $u(x,q)$ equals $(2U(x+q)+U(x-q))/3$, and so the maximal expected utility $ \tilde{u}(x,p)$ at $x=2$ is given by $\max_{q} g(q)$, where
\be
\label{2,p} \textstyle
g(q):=\frac{2}{3}U(2-qp+q)+\frac{1}{3}U(2-qp-q).
\ee 
To find the maximizer $\tilde{q}=\tilde{q}(2,p)$ of $g$, notice that, because of the behavior of $U$ outside $[1,3]$, it is clear that  for small $p$ we only need to consider the interval $[i,s]$ of values of $q$ such that both $2-qp+q$ and $2-qp-q$ belong to $[1,3]$; and since $g$ is affine on this interval, its maximum is attained on the boundary. 
Let us now consider the case of small $p>0$. Simple computations show that $-i=s=(1+p)^{-1}$, and since $U(x)=x-1$ on $[1,3]$ and $(1+p)^{-1}=1-p+o(p)$ we get that
\bd
\textstyle
g(s)=\frac{2}{3}U(2+\frac{1-p}{1+p})+\frac{1}{3}U(2-\frac{1+p}{1+p})=\frac{2}{3}\Big(1+\frac{1-p}{1+p}\Big)=\frac{2}{3}\Big(2-2p+o(p)
\Big).
\ed
and
\bd
\textstyle
g(i)= \frac{2}{3}U(2-\frac{1-p}{1+p})+\frac{1}{3}U(2+\frac{1+p}{1+p})=\frac{2}{3}\Big(1-\frac{1-p}{1+p}\Big)+\frac{2}{3}=\frac{2}{3}\Big(1+2p+o(p)
\Big),
\ed
so that $g(s)>g(i)$. It follows that $$\tilde{q}(2,p)=s=(1+p)^{-1} \quad \text{ and } \quad\tilde{u}(2,p)=g(s)=4/3-4p/3 +o(p).$$

Entirely analogous computations for the case of small $p<0$ give that 
$-i=s=(1-p)^{-1}$ and $g(s)>g(i)$, so that $\tilde{q}(2,p)=s=(1-p)^{-1}=1+p+o(p)$ and
\bd
\textstyle
\tilde{u}(2,p)=g(s)=\frac{2}{3}U(2+\frac{1-p}{1-p})+\frac{1}{3}U(2-\frac{1+p}{1-p})=\frac{4}{3}+ \frac{1}{3}\Big(1-\frac{1+p}{1-p}\Big)=  \frac{2}{3}\Big(2-p +o(p)\Big).
\ed
Putting the pieces together, we obtain that 
\bd
\partial_p\tilde{u}(2,0+)=-\frac{4}{3}<-\frac{2}{3}=\partial_p\tilde{u}(2,0-),
\ed
which shows that $ \tilde{u}(x,\cdot)$ is not convex.

\section{Differentiability}
\label{Differentiability}
In this section we prove the differentiability in $p$ of the value function.
The following differentiability result is trivial, but useful.  \newtheorem{vdiff}[thmerge]{Remark}
 \begin{vdiff}
   \label{vdiff}
   Under the assumptions of Theorem \ref{thmerge} the function
   \begin{eqnarray}
     (0,\infty) \times \mathcal{P} & \longrightarrow & \r \nonumber\\
     (y,p) &\mapsto & \tilde{v}(y,p) \nonumber 
   \end{eqnarray}
   is continuously differentiable, and its derivatives satisfy $\nabla_p \tilde{v}(y,p)= y\nabla_r v(y,yp)$ and  $\partial_y \tilde{v}(y,p)=(\partial_y v + p\nabla_r v)(y,yp)$.
 \end{vdiff}
 \textsf{ PROOF } Since the function $v$ defined in (\ref{dualprrend}) is
 convex and differentiable (see \cite[Lemma 3]{HugonKram:04}) and so
 continuously differentiable (see \cite[Theorem 25.5]{Rock:70}), the
 thesis follows from the identity $\tilde{v}(y,p)=v(y,yp)$.  $\Box$
 \newline

We also record  that  simple explicit examples show that the largest feasible position is not a differentiable function of the derivative's prices.

 To facilitate the proof of Theorem \ref{udiff} we state 
 the following  fact.

 \newtheorem{diflem}[thmerge]{Lemma}
 \begin{diflem}
   \label{diflem}
   Let $f,g,h$ be real valued functions defined on some open set
   $G\subseteq \r^n$, and assume that $f(z)\leq g(z) \leq h(z)$ for
   all $z\in G$, with equality at some $z=\bar{z}\in G$, and assume that $h$ is differentiable at $z=\bar{z}$.
Then, if $g$ is differentiable at $z=\bar{z}$, it has gradient $\nabla
   g(\bar{z})=\nabla h(\bar{z})$. Moreover,  $g$ is differentiable at $z=\bar{z}$ if
   either of the following additional conditions is satisfied:
   \begin{enumerate}
   \item \label{gconv} $g$ is convex.
   \item \label{fdiff}  $f$ is differentiable at $z=\bar{z}$.
   \end{enumerate}
 \end{diflem}
 We forego the proof of the previous real-analysis lemma, as it  is a simple consequence of the squeeze theorem (a.k.a. sandwich theorem) and the following fact: if  $L:\r^n \to \r$ is linear and $L(x)\geq o(x)$ for all $x$ near zero then $L \equiv 0$ (as it follows from  $\pm L( x)=L(\pm \e x)/ \e \geq o( \pm \e x)/\e$, taking limits for $\e \to 0 +$).

We will need the following lemma, a special case of which is the well known  fact that the function $w=u(\cdot,0)$ is differentiable (see \cite[Theorem 2]{KramSch:03}).
We recall that in general the function $u$ is not differentiable on $\K$ (see \cite[Remark 3.1]{HugonKramSch:05}).

 \newtheorem{RadialDiff} [thmerge] {Lemma}
 \begin{RadialDiff}
\label{RadialDiff}
 Under the assumptions of Theorem \ref{thmerge}, the function
 $t \mapsto u(t(x,q)) $ is
     differentiable on $t>0$ if  $(x,q) \in \K$.
\end{RadialDiff} 

\textsf{ PROOF }
Since $(x,q) \in \K$, the thesis follows from \cite[Theorem 23.4]{Rock:70} once we prove that, whenever $(y_i,r_i) \in \partial u(x,q)$, $i=1,2$, the equality $xy_1+qr_1=xy_2+qr_2$ holds.
 The latter follows from the equality $xy_i+qr_i=\mathbb{E}[Y_T(y_i,r_i)(X_T(x,q)+qf)]$, since $Y_T(y_i,r_i)=U'(X_T(x,q)+qf)$ does not depend on $i$ (for the last two identities, see \cite[Theorem 1]{HugonKram:04}).  $\Box$ \newline
 
We remark that the reason why the previous proof does not work when $(x,q) \in \partial \K$ is that the assumptions of \cite[Theorem 23.4]{Rock:70} are not satisfied. The assumptions of this theorem however cannot be weakened: in fact, it is easy to show\footnote{Indeed, consider  the convex function given by $g(x,y):=\max(1-\sqrt{x},|y|)$ for $x\geq 0$, and $g(x,y)=\infty$ if $x<0$; then $\partial g(1,1)=(-\infty,0]\times \{1\}$, yet the function $h(y):=g(0,y)=\max(1,|y|)$ is not differentiable at $y=1$ even if $(a,b)(0,1)$ is constant over $(a,b)\in \partial g(1,1)$.} that the  theorem fails for points not in the (relative) interior of the domain.

The previous lemma allows us to prove differentiability of $ \tilde{u}$ when the solution $(x- \hat{q} p, \hat{q})$ of \eqref{equalus} belongs to $\K$; to show that it does, one can at times use that it equals $-\nabla v(\partial_x \tilde{u} , ( \partial_x \tilde{u})p)$ (see Theorem \ref{comparison}) or that $\tilde{X}_T(x)+\tilde{q}(x)f \geq \e>0$ (see Theorem \ref{th2}) .
Note that in the case of power utilities we can prove differentiability since  somewhat explicit computations are possible; indeed, 
if $U(x)=x^{\alpha} /\alpha$ for some non-zero $\alpha \in
 (-\infty,1)$, then $V(y)=-y^\beta /\beta$ for $\beta:=\alpha / (\alpha
 -1)$.  By homogeneity it easily follows that \bd
 \tilde{v}(y,p)= \frac{-y^\beta }{\beta}
 (-\beta \tilde{v}(1,p)), \ed and since $-\beta \tilde{v}(1,p)>0$ the bi-conjugacy relationships yield 
\be \label{powerut}
\tilde{u}(x,p)=\frac{x^\alpha}{\alpha} (-\beta
 \tilde{v}(1,p))^{1-\alpha} .  
\ee

 \textsf{ PROOF OF THEOREM \ref{udiff}.} 
As explained in
 \cite[Remark 6]{HugonKram:04}, we can assume without loss of generality that
 (\ref{wlog}) holds, so we can use Remark \ref{vdiff}. 
If $U$ is a power utility, the thesis follows from equation (\ref{powerut}).

Denote by 
 $y^\star:=\tilde{y}(x^\star,p^\star), q^\star:=\tilde{q}(x^\star,p^\star)$  the
 optimizers for $p=p^\star$. 
 To prove that $p \mapsto \tilde{u}(x^{\star},p)$ is  differentiable if it is convex, simply apply Lemma \ref{diflem} to the functions $ g(p):=\tilde{u}(x^\star,p), \quad
 h(p):=\tilde{v}(y^{\star},p)+x^\star y^\star$.
 To prove differentiability under the other hypothesis, for $p$ close to $p^\star$ define 
\bd \textstyle
 t(p):=\frac{x^\star}{x^\star+ q^\star(p-p^\star)} \text{, so
   that \,\, } t(p)>0  \text{ and } \,\,  t(p) \big((x^\star-q^\star p^\star, q^\star)(1,p)\big)=x^\star.  \ed
 Taking Lemma \ref{RadialDiff} into account, applying Lemma \ref{diflem}  to the functions
 \bd f(p):=u(t(p) (x^\star-q^\star
 p^\star, q^\star)), \quad g(p):=\tilde{u}(x^\star,p), \quad
 h(p):=\tilde{v}(y^{\star},p)+x^\star y^\star \ed proves that $p \mapsto
 \tilde{u}(x,p)$ is differentiable at $p=p^\star$.
Moreover, whenever $\tilde{u}$ is differentiable,  Lemma \ref{diflem}  shows that
 \bd \nabla_p
 \tilde{u}(x^\star,p^\star)=\nabla_p \tilde{v}(y^\star,p^\star), = y^\star \nabla_r v(y^\star,y^\star p^\star), \ed
which by Theorem \ref{comparison} equals
\bd
\nabla_p \tilde{u}(x^\star,p^\star)= -  q^\star  \partial_x \tilde{u}(x^\star,p^\star).
\ed
 Since  $\tilde{q}$ and $\partial_x \tilde{u}$ are continuous function of $(x,p)$ (by Theorem \ref{continuity1}), the proof is finished.  $\Box$ \newline

 \textbf{Acknowledgements.}  We thank Dmitry Kramkov for his valuable comments on a previous version of this paper, and  Walter Schachermayer for the intuition behind the counter-example in Section \ref{non-convex}.

\end{document}